\documentclass[twocolumn, amsmath, amssymb, aps, notitlepage, nofootinbib,superscriptaddress]{revtex4-1}
\usepackage[normalem]{ulem}
\usepackage{bm,color,float,graphicx}
\usepackage{lipsum}
\usepackage{simplewick}
\usepackage{aas_macros}
\usepackage{natbib}
\usepackage{subcaption}
\usepackage{orcidlink}

\def\m{\mathrm}

\def\h{\hat}

\date{\today}
\begin{document}
\title{Eternal inflation bubble collision signature on CMB remote dipole and quadrupole fields}

\author{Hongbo Cai\orcidlink{0000-0003-3851-7518}}
\email{ketchup@sjtu.edu.cn}
\affiliation{Department of Astronomy, School of Physics and Astronomy, Shanghai Jiao Tong University, Shanghai, 200240, China}
\affiliation{State Key Laboratory of Dark Matter Physics, Shanghai 200240, China}
\affiliation{Key Laboratory for Particle Astrophysics and Cosmology (MOE)/Shanghai Key Laboratory for Particle Physics and Cosmology, Shanghai, China}

\author{Pengjie Zhang\orcidlink{0000-0003-2632-9915}}
\email{zhangpj@sjtu.edu.cn}
\affiliation{Department of Astronomy, School of Physics and Astronomy, Shanghai Jiao Tong University, Shanghai, 200240, China}
\affiliation{Tsung-Dao Lee Institute, Shanghai Jiao Tong University, Shanghai, 200240, China}
\affiliation{State Key Laboratory of Dark Matter Physics, Shanghai 200240, China}
\affiliation{Key Laboratory for Particle Astrophysics and Cosmology (MOE)/Shanghai Key Laboratory for Particle Physics and Cosmology, Shanghai, China}

\author{Yilun Guan\orcidlink{0000-0002-1697-3080}}
\affiliation{Dunlap Institute for Astronomy and Astrophysics, University of Toronto, 50 St. George St., Toronto, ON M5S 3H4, Canada}

\begin{abstract}
The remote dipole and quadrupole fields (RDF/RQF) encode information about the observable universe as seen from remote places within our past light cone. Sensitive to the superhorizon inhomogeneites, they provide a unique way to probe physics at the largest scales, bypassing the limitations of cosmic variance inherent in the primary cosmic microwave background (CMB). In this work, we focus on the bubble collision predicted by the eternal inflation theory, which can leave distinct azimuthally symmetric patterns on the superhorizon scales, potentially detectable through the RDF and RQF. We present the first analytic expression of the RQF signal induced by bubble collision and validate it against numerical calculations performed with \texttt{RemoteField}, a new public software tool we developed, finding excellent agreement between the two. Combining our new RQF calculation with the corresponding RDF signal calculated by prior work, we forecast the constraining power on bubble collision parameters using RDF/RQF reconstruction.
We find that, for an CMB-S4-like and an LSST-like experiment, the RDF reconstruction can provide comparable constraining power as that from the primary CMB alone; and the RQF reconstruction can improve the constraining power by about an order of magnitude. We argue that these constraints can be improved further by including more RDF/RQF multipoles and by using tomographic techniques to mitigate the standard $\Lambda$CDM signal. We anticipate the framework we developed in this work to be broadly applicable to probe other superhorizon-scale physics, such as cosmic topology and domain walls.
\end{abstract}
\maketitle

\section{Introduction}
\label{sec:introduction}
The anisotropies of the primary cosmic microwave background (CMB) provide a powerful and direct probe of early-universe physics, as they are primarily sourced by photons scattered near the last scattering surface (LSS). Large angular scale CMB modes naturally encode primordial universe information and can constrain physics beyond the standard cosmological model, revealing features such as a hemi-spherical power asymmetry and a lack of correlations on large angular scales (for review, see \cite{Schwarz:2015cma}). However, measurements of large-scale primary CMB by experiments like Planck \cite{Planck2018:VI:CP}, ACT \cite{ACT:2025fju}, SPT \cite{SPT-3G:2025bzu}, and BICEP \cite{Chiang:2009xsa} are fundamentally limited by cosmic variance limit, preventing higher statistical significance for these anomalies.

Secondary CMB anisotropies offer a powerful alternative to probe superhorizon-scale physics, effectively bypassing the cosmic variance limit of the primary CMB. The most prominent of these are the kinetic and polarized Sunyaev-Zel'dovich effects (kSZ \cite{Sunyaev:1980nv,1987ApJ...322..597V,Zhang:2003nr}; pSZ \cite{Sazonov:1999zp,Deutsch:2017cja}). These effects arise from Thomson scattering of CMB photons by the bulk motions of ionized electrons along the line of sight, producing temperature and polarization anisotropies that dominant the small-scale (arcminute-scale) CMB power spectrum. While the kSZ effect has been detected robustly \cite{Hand_2012,DeBernardis:2016pdv,Chen:2021pwg,Kusiak:2021hai}, the pSZ remains elusive but is anticipated to be measured with upcoming experiments like the Simons Observatory and CMB-S4.

The key fields for the kSZ/pSZ effects are the remote dipole field (RDF; also known as the effective radial velocity field) and remote quadrupole field (RQF), defined as the effective dipole and quadrupole moments of CMB temperature anisotropies in the local electron's rest frame \cite{Kamionkowski:1997na}. The kSZ and pSZ effects arise from the coupling of the ionized electron overdensity with the RDF/RQF along the line of sight, which imprint secondary temperature and polarization anisotropies on the small-scale CMB\footnote{The coupling between RQF and the mean density of ionized electrons produces a low-$\ell$ polarization signal known as the ``reionization bump''.} (see \cite{Deutsch:2018, Deutsch:2017cja} for reviews). 

Unlike the primary CMB, which is cosmic-variance limited at large angular scales, RDF/RQF are sensitive to superhorizon inhomogeneities and have been proposed to probe the CMB anomalies \cite{Cayuso:2019hen}, cosmic birefringence \cite{Namikawa:2023zux}, local non-Gaussianities \cite{Munchmeyer:2018eey}, and primordial gravitational waves \cite{Deutsch:2018umo}. In this work, we
focus on bubble collision predicted by the eternal inflation model \cite{Garriga:2006hw, Kosowsky:1991ua}, a theoretical compelling but observationally unconfirmed phenomenon. In this scenario, our observable universe resides within one of many bubbles nucleated during eternal inflation, driven by a false vacuum, as a tractable realization of the multiverse. Collisions between bubbles imprints relic superhorizon-scale\footnote{In this paper, we use “horizon” to denote the present-day Hubble radius, which is simply $(H_{0})^{-1}$ with $c=1$.} inhomogeneities \cite{Wainwright:2014pta}, generating azimuthally symmetric patterns in the RDF/RQF. This motivates the use of RDF/RQF to constrain the bubble collision scenerio, necessitating a careful modeling of the induced signals. Building on prior work on the RDF \cite{Zhang:2015uta}, we derive the first analytic expression for the RQF signal induced by the bubble collision and validate it against numerical results. The numerical results are performed using \texttt{RemoteField},\footnote{\url{https://github.com/catketchup/RemoteField}} a publicly available software tool we developed for RDF/RQF numerical calculation applicable to a wide range of superhorizon-scale phenomena.

The RDF and RQF can be reconstructed on our past light cone by combining CMB experiments with galaxy surveys using a quadratic estimator \cite{Deutsch:2018},
leveraging the coupling between CMB photons and electrons on small angular scales to recover large-scale RDF/RQF signals.
This method has been demonstrated in practice in a joint Planck and unWise analysis \cite{Bloch:2024} and a joint analysis of data from ACT and DESI LRGs \cite{McCarthy:2024nik}. It has also been applied to constrain on other phenomena such as Gpc-scale voids \cite{PhysRevLett.107.041301}, the intrinsic dipole, and matter-radiation isocurvature modes \cite{Krywonos:2024mpb}. Building on this framework, in this work we forecast the constraining power of RDF and RQF reconstruction on bubble collision scenario.

The paper is organized as follows. Section~\ref{sec:kSZ/pSZ and RDF/RQF} reviews the kSZ/pSZ effects and the RDF/RQF. Section~\ref{sec:BC signal on RQF} demonstrates the analytic calculation of the RQF induced by the bubble collision from the eternal inflation and validates it with numerical results. Section~\ref{sec:forecast} reviews the RDF/RQF reconstruction method and presents the forecasted constraints on the bubble collision parameters using the RDF/RQF reconstruction. We conclude in Section~\ref{sec:conclusion}.

\section{The polarized Sunyaev Zel’dovich effect and the remote quadrupole field}
\label{sec:kSZ/pSZ and RDF/RQF}
In this section, we review the kSZ/pSZ effects which share similar mathematical formalism, and introduce the RDF/RQF as the key elements for kSZ/pSZ effects. We then present the numerical method with Fourier kernels.

\subsection{kSZ/pSZ effects and RDF/RQF}
The kinetic and polarized Sunyaev-Zel'dovich (kSZ/pSZ) effects are secondary anisotropies in the CMB. They arise when CMB photons scatter off free electrons moving with a bulk velocity relative to the CMB rest frame. This process transfers power from the local radiation field at the scattering location to the CMB we observe today. These effects are sourced by the coupling between the electron distribution and the local CMB anisotropy field, integrated along the line of sight, $\hat{\bm{n}}_e$. The contributions of kSZ and pSZ to the CMB temperature ($\Theta$) and polarization fields ($Q/U$) are given by
\begin{equation}
  \label{eq:kSZ pSZ LOS}
  \begin{aligned}
    \Theta^{\mathrm{kSZ}}(\hat{\bm{n}}_e) &= \int d \chi_e \dot{\tau}(\hat{\bm{n}}_e, \chi_e) v(\hat{\mathbf{n}}_e, \chi_e),\\
    (Q \pm i U)^{\mathrm{pSZ}}(\hat{\bm{n}}_e) &=  \frac{\sqrt{6}}{10} \int d \chi_e \dot{\tau}(\hat{\bm{n}}_e, \chi_e) q^{\pm}(\hat{\mathbf{n}}_e, \chi_e),
    \end{aligned}
\end{equation}
where $\chi_e$ is the comoving radial distance to the electron from us. The key quantities are the remote dipole field (RDF), $v(\hat{\mathbf{n}}_e, \chi_e)$, and the remote quadrople field (RQF), $q^{\pm}(\hat{\mathbf{n}}_e, \chi_e)$. $\dot{\tau}(\hat{\bm{n}}_e, \chi_e)$ is the differential optical depth, which depends on the local density of free electrons,
\begin{equation}
\dot{\tau}(\hat{\bm{n}}_e, \chi_e) = -\sigma_{T} a_e (\chi_e)\bar{n}_e(\chi_e) (1+\delta_e(\hat{\bm{n}}_e, \chi_e)),
\end{equation}
where $\sigma_T$ is the Thomson scattering cross-section, $a_e(\chi_e)$ is the scale factor, $\bar{n}_e(\chi_e)$ and $\delta_e(\hat{\bm{n}}_e, \chi_e)$ are the average electron number density and the corresponding fractional overdensity, respectively.

The RDF and RQF can be decomposed as
\begin{equation}
  \label{eq:RDF and RQF}
  \begin{aligned}
      v(\hat{\mathbf{n}}_e, \chi_e)
& \equiv \sum_{m=-1}^{1} v^{m}(\hat{\mathbf{n}}_e, \chi_e) Y_{1m}(\hat{\mathbf{n}}_e), \\
q^{\pm}(\hat{\mathbf{n}}_e, \chi_e)
& \equiv \sum_{m=-2}^{2} q^{m}(\hat{\mathbf{n}}_e, \chi_e) {}_{ \pm 2}Y_{2m}(\hat{\mathbf{n}}_e),
    \end{aligned}
\end{equation}
where $v^m(\hat{\bm{n}}_e, \chi_e)$ and $q^m(\hat{\bm{n}}_e, \chi_e)$ are the CMB temperature dipole and polarization quadrupole multipoles observed by an electron at the comoving coordinate position $(\hat{\bm{n}}_e, \chi_e)$. In a certain position $(\hat{\bm{n}}_e, \chi_e)$, $v^m(\hat{\bm{n}}_e, \chi_e)$ and $q^m(\hat{\bm{n}}_e, \chi_e)$ should be calculated by
\begin{equation}
  \label{eq:veff and qeff}
  \begin{aligned}
        v^m(\hat{\bm{n}}_e, \chi_e)&=\int_{\Omega} d^2 \hat{\bm{n}}\,  \Theta(\chi_e, \hat{\bm{n}}_e, \hat{\bm{n}}) Y_{1 m}^*(\hat{\bm{n}}),\\
    q^m(\hat{\bm{n}}_e, \chi_e)&=\int_{\Omega} d^2 \hat{\bm{n}}\,  \Theta(\chi_e, \hat{\bm{n}}_e, \hat{\bm{n}}) Y_{2 m}^*(\hat{\bm{n}}),
    \end{aligned}
  \end{equation}
where $\hat{\bm{n}}$ denotes the direction vector in the local electron coordinate and $\Theta(\chi_e, \hat{\bm{n}}_e, \hat{\bm{n}})$ is the CMB temperature anisotropies in the electron LSS.

As Eq.~\eqref{eq:RDF and RQF} demonstrate, RDF/RQF can be seen as the effective projection of the local dipole/quadrupole moments to the direction of the line of sight; the former is a spin-0 field while the latter is a spin-2 field.

One can decompose $q^{\pm}$ with spin-weighted spherical harmonics to obtain rotation-invariant quantities, as \cite{Deutsch:2017cja, Deutsch:2018}
\begin{equation}
  \label{eq:RQF multipole}
q_{\ell m}^{\pm}(\chi_{e}) = \int_{\Omega} d^{2}\hat{\mathbf{n}}_{e} \,
q^{\pm}(\hat{\mathbf{n}}_{e}, \chi_{e}) \,
{}_{\pm 2}Y_{\ell m}^{*}(\hat{\mathbf{n}}_{e}),
\end{equation}
and get the E and B-mode of RQF by
\begin{equation}
  \label{eq: RQF EB mode}
  \begin{aligned}
q^{E}_{\ell m} &= -\frac{1}{2} ( q^{+}_{\ell m} + q^{-}_{\ell m} ), \\
q^{B}_{\ell m} &= -\frac{1}{2\mathrm{i}} ( q^{+}_{\ell m} - q^{-}_{\ell m} ).
  \end{aligned}
\end{equation}
Note that if the RQF is sourced only by scalar perturbations, then $q_{\ell m}^{+}= q_{\ell m}^{-}$ and $q^{B}_{\ell m}=0$ \cite{Deutsch:2018}.\footnote{Under this circumstance, the B-mode of the pSZ effect still exists due to the coupling of the optical depth field fluctuations with the RQF.} This is the case we discuss in this work, that is, to investigate the RDF/RQF contributed by the additional primordial scalar perturbations contributed by the bubble collision.


\subsection{Local CMB temperature anisotropies decomposition}
\label{subsec:fourier kernels}
One can decompose the local CMB temperature anisotropies $\Theta(\chi_e, \hat{\bm{n}}_e, \hat{\bm{n}})$ as
\begin{equation}
  \label{eq:CMB contributions}
  \begin{aligned}
    \Theta(\chi_e, \hat{\bm{n}}_e, \hat{\bm{n}}) = \Theta_{\mathrm{SW}}(\chi_e, \hat{\bm{n}}_e, \hat{\bm{n}}) &+ \Theta_{\mathrm{ISW}}(\chi_e, \hat{\bm{n}}_e, \hat{\bm{n}})  \\
    &+ \Theta_{\mathrm{Doppler}}(\chi_e, \hat{\bm{n}}_e, \hat{\bm{n}}),
    \end{aligned}
\end{equation}
where the three contributions represent the Sachs-Wolfe (SW) effect generated by the gravitational potential difference from the last-scattering surface, the integrated Sachs-Wolfe (ISW) effect from the temporal evolution of gravitational potential, and the Doppler effect due to both the peculiar motion of the electrons emitted at the last scattering surface and the peculiar motion of the local free-falling electron relative to the CMB rest frame.\footnote{Following \cite{Meerburg:2017xga}, the CMB rest frame is defined as the one in which
the locally observed aberration effect vanishes.}

In the Newtonian gauge, the perturbed FRW universe (neglecting spatial curvature) is given by
\begin{equation}
  \label{eq:metric}
  ds^{2} = -(1 + 2\Psi)dt^{2} + a^{2}(t)(1 - 2\Psi)d\mathbf{x}^{2},
\end{equation}
in the matter and dark energy dominated era when the anisotropic stress tensor is approximately zero, and $\Psi$ is the gravitational potential at the time. On large scales, the three contributions in Eq.~\eqref{eq:CMB contributions} can be expressed as functions of primordial scalar perturbations $\Psi_i$ \citep{Erickcek:2008jp},
\begin{equation}
  \label{eq:3 contributions}
\begin{aligned}
\Theta_{\mathrm{SW}}(\hat{\bm{n}}_e, \chi_e, \hat{\bm{n}}) & =(2 D_{\Psi}(\chi_{\mathrm{dec}})-\frac{3}{2}) \Psi_i(\bm{r}_{\mathrm{dec}}), \\
\Theta_{\mathrm{ISW}}(\hat{\bm{n}}_e, \chi_e, \hat{\bm{n}}) & =2 \int_{a_{\mathrm{dec}}}^{a_e} \frac{d D_{\Psi}}{d a} \Psi_i(\bm{r}(a)) d a, \\
  \Theta_{\mathrm{Doppler}}(\hat{\bm{n}}_e, \chi_e, \hat{\bm{n}}) &= \hat{\bm{n}} \cdot\bigl[D_v(\chi_{\mathrm{dec}}) \nabla \Psi_i(\bm{r}_{\mathrm{dec}}) \\&
  -D_v(\chi_e) \nabla \Psi_i(\bm{r}_e)\bigr],
\end{aligned}
\end{equation}
where $D_{\Psi}(\chi_{\mathrm{dec}})$ is the potential growth function, and $D_{v}(\chi_{\mathrm{dec}})$ is the velocity growth function (see Appendix~\ref{sec:appendix} for details).
$\bm{r}_{\text{dec}} \equiv \chi _ { e } \hat { \mathbf{n}} _ { e } + \Delta \chi_{\text{dec}}  \hat {\mathbf{n}}$ is the position vector of decoupling, $\chi_{\text{dec}}$ is the total distance from us to decoupling, and $\Delta \chi_{\text{dec}} \equiv -\int_{a_{\text{e}}}^{a_{\text{dec}}} da\Bigl( H(a)a^2 \Bigr)^{-1}$ is the distance from the electron to decoupling.

\subsection{RDF/RQF numerical calculation}
Relating the three contributions in Eq.~\eqref{eq:3 contributions} with a given primordial potential $\Psi_i$, we are able to numerically calculate the RDF/RQF as \cite{Terrana:2016xvc, Deutsch:2017cja}
\\
\begin{widetext}
  \begin{equation}
    \label{eq:RDF/RQF Fourier}
    \begin{aligned}
      v(\hat{\mathbf{n}}_e, \chi_e) & = \sum_{m=-1}^{1} \biggl[\int \frac{d^3\bm{k}}{(2\pi)^3} \, \Psi_i(\bm{k}) T(k)\mathcal{K}(k, \chi_e)Y_{1m}^*(\hat{\bm{k}}) \,e^{\mathrm{i} \chi_e \bm{k} \cdot \hat{\mathbf{n}}_e}\biggr] Y_{1m}(\hat{\mathbf{n}}_e), \\
      q^{\pm}(\hat{\mathbf{n}}_e, \chi_e) &= \sum_{m=-2}^{2} \biggl[\int \frac{d^3\bm{k}}{(2\pi)^3} \, \Psi_i(\bm{k}) T(k)\mathcal{G}(k, \chi_e)Y_{2m}^*(\hat{\bm{k}}) \,e^{\mathrm{i} \chi_e \bm{k} \cdot \hat{\mathbf{n}}_e}\biggr] {}_{ \pm 2}Y_{2m}(\hat{\mathbf{n}}_e),
    \end{aligned}
\end{equation}

\end{widetext}
where $T(k)$ is the transfer function accounting for sub-horizon evolution on small scales, $\mathcal{K}$ and $\mathcal{G}$ are the Fourier kernels for RDF and RQF consisting of three contributions corresponding to the counterparts in Eq.~\eqref{eq:CMB contributions} given by
\begin{equation}
  \label{eq:G and K components}
      \begin{aligned}
        \mathcal{K}(k,\chi_e) &\equiv \mathcal{K}_{\mathrm{SW}}(k, \chi_e) + \mathcal{K}_{\mathrm{ISW}}(k, \chi_e) + \mathcal{K}_{\mathrm{Doppler}}(k, \chi_e),\\
        \mathcal{G}(k,\chi_e) &\equiv \mathcal{G}_{\mathrm{SW}}(k, \chi_e) + \mathcal{G}_{\mathrm{ISW}}(k, \chi_e) + \mathcal{G}_{\mathrm{Doppler}}(k, \chi_e).
        \end{aligned}
\end{equation}
For detailed expressions of the above transfer function and kernels see Appendex~\ref{sec:appendix}. We develop a software \texttt{RemoteField} which numerically calculate the RDF/RQF using Eq.~\eqref{eq:RDF/RQF Fourier}.
The code takes a primordial potential, $\Psi_i(\bm{k})$, as the input and allows for custom transfer function, $T(k)$. This code is implemented based on kernels provided by \texttt{SZ\_cosmo},\footnote{\url{https://github.com/rcayuso/SZ_cosmo}} and utilizes FFT algorithm to perform efficient calculation of a 3D RDF/RQF distribution at a given cosmic time. The 2D spherical distribution at a given comoving distance can then be obtained by interpolating the 3D RDF/RQF on a grid. We use this code to calculate numerically the RQF induced by the bubble collision as a cross-check with the analytic result that will be discussed in the subsequent section.

\section{Bubble collision signal on remote quadrupole field}
\label{sec:BC signal on RQF}
In this section, we calculate the RQF signal contributed by the eternal inflation bubble collision. We follow a similar method used to calculate the corresponding RDF signal given by \cite{Zhang:2015uta}.

\subsection{The spacetime in the aftermath of a bubble collision}
\begin{figure*}[t]
  \includegraphics[width=\textwidth]{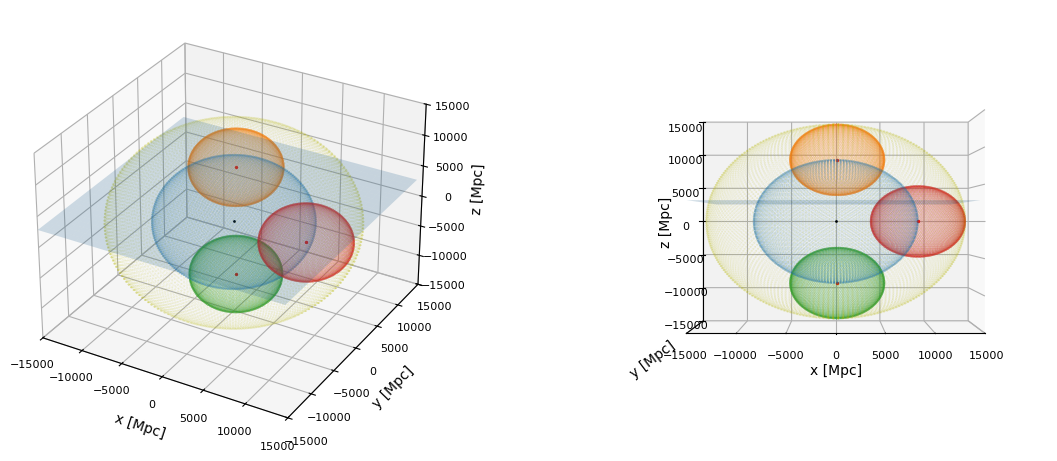}
  \caption{Schematic diagrams showing different regions in the comoving space relevant to the bubble collision with $Z_c = 0.8$ from two different angles of view. The large yellow sphere and the large blue sphere depict our LSS and the electrons at $Z_e=7$ respectively. The flat surface represents the boundary with the upper region affected by the bubble collision and the lower region unaffected. The small spheres denote the electron LSS of three typical cases: entirely outside the region affected by the bubble collision (green), intersecting with the boundary (red), and lying entirely with in the region affected by the bubble collision (orange).}
  \label{fig:schematic bubbles}
\end{figure*}

The spatial distribution of the primordial potential induced by the collision between two bubbles, $\Psi^{\text{bc}}_i$, follows a SO(2,1) symmetry and can be described approximately by a linear and a quadratic term as \cite{Wainwright:2014pta},
\begin{equation}
  \label{eq:BC potential}
\Psi^{\text{bc}}_i(\mathbf{r}) \sim
\begin{cases}
A(\frac{x}{r_{H}} - \frac{x_c}{r_H}) + B(\frac{x}{r_{H}} - \frac{x_c}{r_H})^2, & \text{if } x \geq x_c, \\
0, & \text{if } x < x_c,
\end{cases}
\end{equation}
where $r_H \equiv \frac{1}{H_0}$ is the present Hubble radius \footnote{In the overall paper, we take the natural unit $c=1$ as the speed of light.},  $x_c$ is the location of the causal boundary, and A and B are coefficients related to the fundamental parameters of the eternal inflation (see Appendix~\ref{sec:appendix b}). The causal boundary can be treated as flat over the size of the present day horizon, i.e., left by two much larger colliding bubbles. For convenience, we fix the z-axis of the coordinate along the direction of the bubble collision, and a schematic diagram is shown in Fig.~\ref{fig:schematic bubbles} with $Z_c = 0.8$ where $Z_c$ is the closest redshift to the causal boundary of the collision. In the diagram, the blue flat surface represents the boundary splitting the upper region affected by the bubble collision and the lower region unaffected.
The largest yellow sphere represents our LSS, and the smaller blue sphere denotes a redshift slice at $Z_e=7$. The three small spheres (red, green, and orange) represent the LSS's of the local electrons at the redshift slice $Z_e=7$. We depict three typical cases: (1) the electron's LSS lies outside the bubble collision region (green), (2) the electron's LSS lies entirely within the bubble collision-affected region (red), and (3) the LSS is only partially within the bubble collision-affected region (orange).

\subsection{Analytical calculation of the RQF signal induced by the bubble collision}
\label{subsec:analytical calculation}
Using Eq.~\eqref{eq:RDF and RQF} and Eq.~\eqref{eq:veff and qeff}, we derive the analytical form of the RQF induced by the bubble collision. While the general method requires calculating all moments of  $q^{m}(\hat{\mathbf{n}}_e, \chi_e)$ and pairing them with spin-weighted spherical harmonics ${}_{ \pm 2}Y_{2m}$, the SO(2,1) symmetry of bubble collision induced perturbation allows for a simplification. By aligning the z-axis with the collision direction, the resulting CMB temperature in the local electron frame,
$\Theta_{\mathrm{\alpha}}(\chi_e, \hat{\bm{n}}_e, \hat{\bm{n}})$ $(\alpha \in \{ \mathrm{SW, ISW, Doppler}\})$, becomes axially symmetric. This symmetry ensures that $q^m$ vanishes except when $m=0$. We therefore compute only $q_{\mathrm{ \alpha}}^{0}$, yielding an RQF that depends only on the polar angle $\theta$ with respect to the positive z-axis. We omit the $\pm$ superscript as ${}_{2}Y_{20} = {}_{-2}Y_{20}$, and the RQF field is given by
\begin{equation}
  \label{eq:q_alpha}
  \begin{aligned}
q_{\mathrm{\alpha}}(\hat{\bm{n}}_e, \chi_e)  &= q_{\mathrm{ \alpha}}^{0}(\hat{\bm{n}}_e, \chi_e) {}_{2}Y_{2 0}(\hat{\bm{n}}_{e}) \\ &= (\int_{\Omega} d^2 \hat{\bm{n}}\;\Theta_{\mathrm{\alpha}}(\chi_e, \hat{\bm{n}}_e, \hat{\bm{n}})Y_{2 0}(\hat{\bm{n}})) {}_{2}Y_{2 0}(\hat{\bm{n}}_{e}).
  \end{aligned}
\end{equation}
Given the axially-symmetry of $\Theta_{\mathrm{\alpha}}(\chi_e, \hat{\bm{n}}_e, \hat{\bm{n}})$ $(\alpha \in \{ \mathrm{SW, ISW, Doppler}\})$, we can further simplify the integration by
\begin{equation}
    \label{}
      \begin{aligned}
        q_{\alpha}^{0}(\hat{\bm{n}}_e, \chi_e) & = \int_{\Omega} d^2 \hat{\bm{n}}\; \Theta_{\mathrm{\alpha}}(\chi_e, \hat{\bm{n}}_e, \hat{\bm{n}})
                                                                      Y_{2 0}(\hat{\bm{n}}) \\ & = 2\pi \int_{0}^{\theta_{\m{c}}} d \theta\;\sin\theta (3\cos^2\theta - 1) \\ & \,\,\,\,\,\,  \times \Theta_{\mathrm{\alpha}}(\chi_e, \hat{\bm{n}}_e, \theta),
\end{aligned}
\end{equation}
The integration bounds over $\theta$ depends the relative position of the local electron's LSS with respect to the bubble collision surface. As depicted in Fig.~\ref{fig:schematic bubbles}, when the electron's LSS lies outside, within, and partially within the bubble collision boundary, the upper limit, $\theta_c$, is set to $0$, $\pi$, and $\cos \theta_c = (x_{\m{c}} - \chi_{\m{e}}\cos\theta_{\m{e}})/\Delta\chi_{\m{dec}}$, respectively.\footnote{When the LSS lies entirely outside the region affected by the bubble collision, the corresponding RQF signal equals zero naturally. For consistency in form, we can set $\theta_c=0$.} It is also convenient to express $\Psi^{\text{bc}}_i$ and its gradient in the spherical coordinates as
\begin{equation}
  \label{}
  \begin{aligned}
   \Psi^{\text{bc}}_i(\bm{r}_{\mathrm{dec}}) &= \frac{A}{r_{\m{H}}} ((\chi_{\m{e}}\cos\theta_{e} + \Delta \chi_{\m{dec}}) \cos\theta - x_{\m{c}} )  \\ &+ \frac{B}{r^2_{\m{H}}}((\chi_{\m{e}}\cos\theta_{e} + \Delta \chi_{\m{dec}})\cos\theta - x_{\m{c}} )^{2},
  \end{aligned}
 \end{equation}
 and
\begin{equation}
  \label{}
 \nabla \Psi^{\text{bc}}_i(\bm{r}_{\mathrm{dec}}) = \frac{A}{r_{\m{H}}} + \frac{2B}{r^2_{\m{H}}}((\chi_{\m{e}}\cos\theta_{e} + \Delta \chi_{\m{dec}})\cos\theta - x_{\m{c}} ).
 \end{equation}

 \begin{figure*}[t]
   \includegraphics[width=0.9\textwidth]{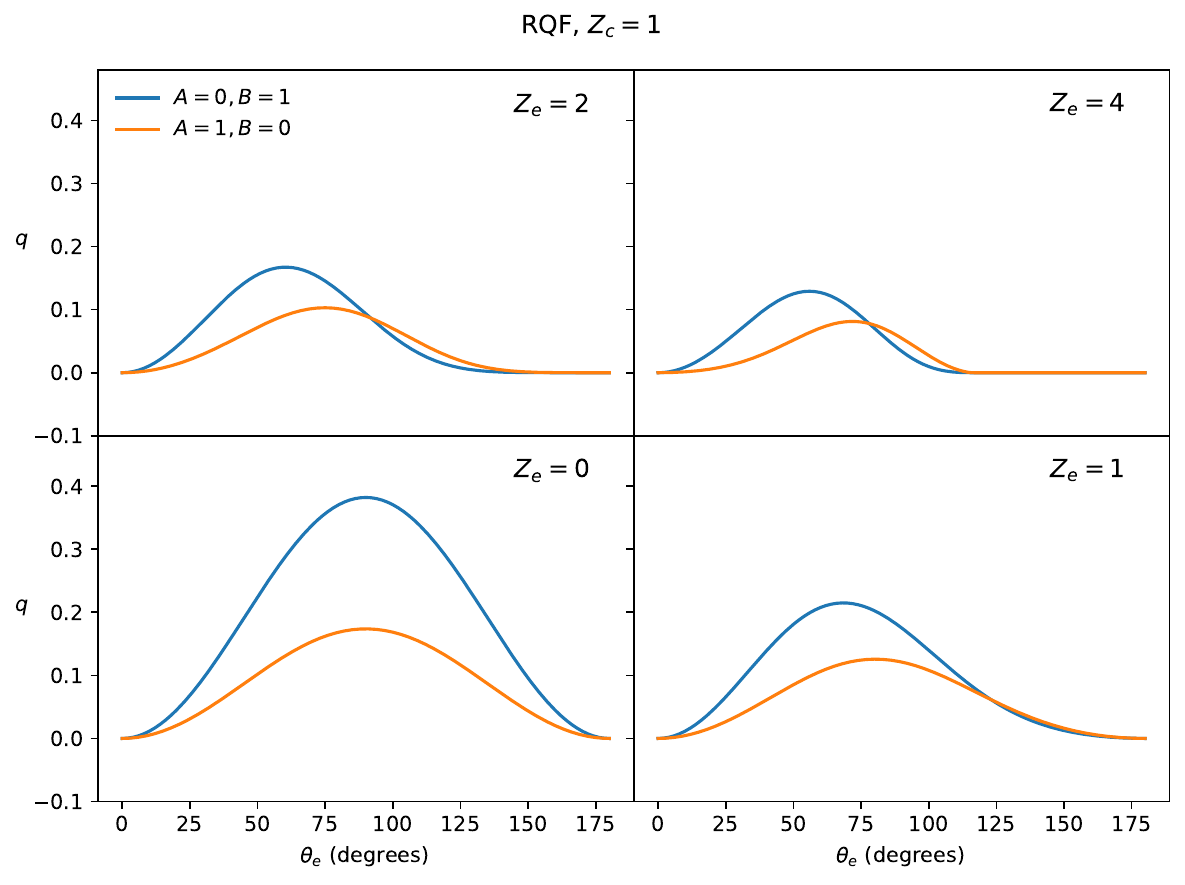}
   \caption{The remote quadrupole field induced by the bubble collision with $Z_{c}=1$. In a coordinate that has the z-axis aligned with the direction of the collision, we plot the 1-D RQFs observed by the electrons at $Z_e = 0, 1, 2, 4$ as functions $\theta_{e}$. The total RQF is the sum of the SW, Doppler, and ISW contributions as given in Eq.~\eqref{eq:q_SW}, Eq.~\eqref{eq:q_Doppler}, and Eq.~\eqref{eq:q_ISW}, respectively. The orange curves are for $A=1, B=0$, and the blue curves are for $B=1, A=0$. The RQF approaches zero when $\theta_{e}\rightarrow 0$ and $\theta_{e}\rightarrow \pi$ because of the spin-weighted spherical harmonics ${}_{2}Y_{2 0}$. $q$ is dimensionless, as a fraction of the speed of light $c$.}
   \label{fig:RQF zc=1}
\end{figure*}


\begin{figure*}[t]
  \includegraphics[width=0.9\textwidth]{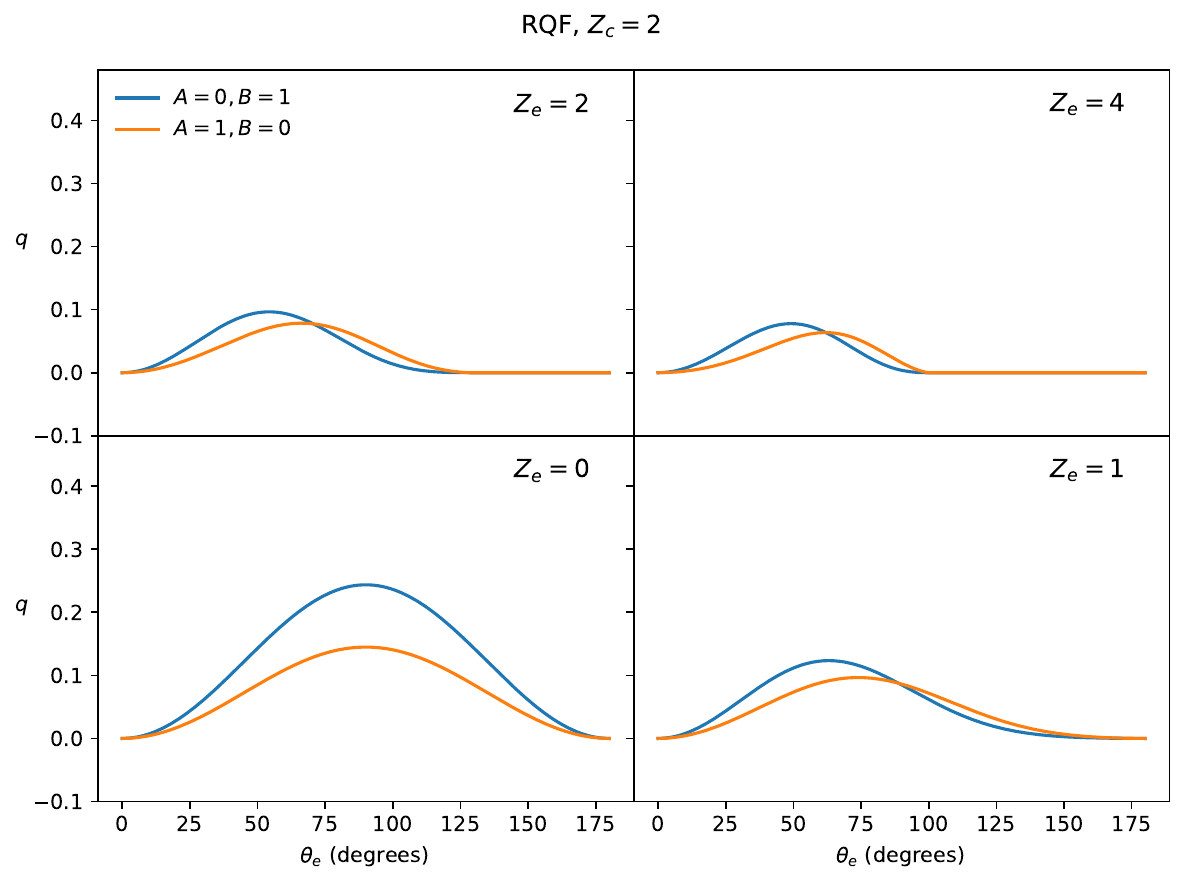}
  \caption{Same as Fig.~\ref{fig:RQF zc=1}, the remote quadrupole field caused by the bubble collision for $Z_{c}=2$. The orange curves are for $A=1, B=0$ , and the blue curves are for $B=1, A=0$.}
    \label{fig:RQF zc=2}
\end{figure*}

Next we calculate the RQF signal of the three contributions from the bubble collision by plugging $\Theta_{\alpha}$ into Eq.~\eqref{eq:q_alpha} $(\alpha \in \{ \mathrm{SW, ISW, Doppler}\})$. $q_{\mathrm{SW}}$ and $q_{\mathrm{Doppler}}$ can be expressed analytically as
\begin{widetext}
\begin{equation}
\label{eq:q_SW}
\begin{aligned}
   q_{\text{SW}}(\hat{\bm{n}}_e, \chi_e) &=
                                                                                                         (2 D_{\Psi}(\chi_{\mathrm{dec}})-\frac{3}{2}) \frac{5\sqrt{6}}{16} \sin^2 \theta_{e}\\ &
                                                                                                        \times \Biggl\{ \frac{A}{r_{\m{H}}}\biggl[\frac{3}{4}\Delta\chi_{\m{dec}}\cos^4\theta + (\chi_{\m{e}}\cos\theta_{\m{e}}-x_{\m{c}})\cos^3\theta - \frac{1}{2}\Delta\chi_{\m{dec}}\cos^2\theta - (\chi_{\m{e}}\cos\theta_{\m{e}}-x_{\m{c}})\cos\theta \biggr] \\ & + \frac{B}{r^2_{\m{H}}} \biggl[\frac{3}{5} \Delta\chi_{\m{dec}}^2 \cos^5\theta + \frac{3}{2}\Delta\chi_{\m{dec}}(\chi_{\m{e}}\cos\theta_{\m{e}}-x_{\m{c}})\cos^4\theta + \frac{1}{3}[-\Delta\chi_{\m{dec}}^2+3(\chi_{\m{e}}\cos\theta_{\m{e}}-x_{\m{c}})^2]\cos^3\theta\ \\ &  - \Delta\chi_{\m{dec}} (\chi_{\m{e}}\cos\theta_{\m{e}}-x_{\m{c}})\cos^2\theta -(\chi_{\m{e}}\cos\theta_{\m{e}}-x_{\m{c}})^2\cos\theta\biggr] \Biggr\}^{\theta=0}_{\theta=\theta_{\m{c}}},
\end{aligned}
\end{equation}

\begin{equation}
\label{eq:q_Doppler}
\begin{aligned}
  q_{\mathrm{ Doppler}}(\hat{\bm{n}}_e, \chi_e) & = D_{v}(\chi_{\mathrm{dec}}) \frac{5\sqrt{6}}{16} \sin^2 \theta_{e} \\ & \times \Biggl\{\frac{A}{r_{\m{H}}} \biggl[\frac{3}{4}\cos^4\theta - \frac{1}{2} \cos^2\theta \biggr]  \\ & + \frac{2B}{{r_{\m{H}}}^2} \biggl[\frac{3}{5}\Delta\chi_{\m{dec}}\cos^5\theta + \frac{3}{4 }(\chi_{\m{e}}\cos\theta_{\m{e}}-x_{\m{c}})\cos^4\theta - \frac{1}{3}\Delta\chi_{\m{dec}}\cos^3\theta  - \frac{1}{2}(\chi_{\m{e}}\cos\theta_{\m{e}}-x_{\m{c}})\cos^2\theta \biggr] \Biggr\}^{\theta=0}_{\theta=\theta_{\m{c}}}.
\end{aligned}
\end{equation}
\end{widetext}
Note that the local Doppler contribution, $D_v(\chi_e) \nabla \Psi_i(\bm{r}_e)$, which is induced by the peculiar velocity of the electron frame in $\Theta_{\mathrm{Doppler}}$ (in Eq.~\eqref{eq:3 contributions}) does not contribute to the RQF as it generates a pure dipole signal to the leading order.

$q_{\mathrm{ ISW}}$ is given by integrating the SW contribution over the line-of-sight as
\begin{equation}
  \label{eq:q_ISW}
  \begin{aligned}
    & q_{\mathrm{ ISW}}(\hat{\bm{n}}_e, \chi_e) \\= &2 \int_{a_{\mathrm{dec}}}^{a_e} \frac{d D_{\Psi}}{d a}\Bigl(\int_{\Omega} d^2 \hat{\bm{n}}\, \Psi_i(\bm{r}(a)) Y_{2 0}(\hat{\bm{n}} ) \Bigr) \, {}_{2}Y_{2 0}(\hat{\bm{n}}_{e}) da,
  \end{aligned}
\end{equation}
\begin{widetext}
where
\begin{equation}
  \label{}
\begin{aligned}
&\Bigl(\int_{\Omega} d^2 \hat{\bm{n}}\, \Psi_i(\bm{r}(a)) Y_{2 0}(\hat{\bm{n}} ) \Bigr) \, {}_{2}Y_{2 0}(\hat{\bm{n}}_{e}) \\ = & \frac{5\sqrt{6}}{16} \sin^2 \theta_{e}
                                                                                     \times \Biggl\{ \frac{A}{r_{\m{H}}}\biggl[\frac{3}{4}\Delta\chi(a)\cos^4\theta + (\chi_{\m{e}}\cos\theta_{\m{e}}-x_{\m{c}})\cos^3\theta - \frac{1}{2}\Delta\chi(a)\cos^2\theta - (\chi_{\m{e}}\cos\theta_{\m{e}}-x_{\m{c}})\cos\theta \biggr] \\ &  + \frac{B}{r^2_{\m{H}}} \biggl[\frac{3}{5} \Delta\chi(a)^2 \cos^5\theta + \frac{3}{2} \Delta\chi(a) (\chi_{\m{e}}\cos\theta_{\m{e}}-x_{\m{c}})\cos^4\theta + \frac{1}{3}[-\Delta\chi(a)^2+3(\chi_{\m{e}}\cos\theta_{\m{e}}-x_{\m{c}})^2]\cos^3\theta\ \\ & - \Delta\chi(a) (\chi_{\m{e}}\cos\theta_{\m{e}}-x_{\m{c}})\cos^2\theta -(\chi_{\m{e}}\cos\theta_{\m{e}}-x_{\m{c}})^2 \cos\theta\biggr] \Biggr\}^{\theta=0}_{\theta=\theta_{\m{c}}(a)},
\end{aligned}
\end{equation}
\end{widetext}
where $\theta_{\m{c}}(a)$ is calculated like $\theta_{\m{c}}$ but as a function of $a$.





\begin{figure*}[t]
  \includegraphics[width=0.6\textwidth]{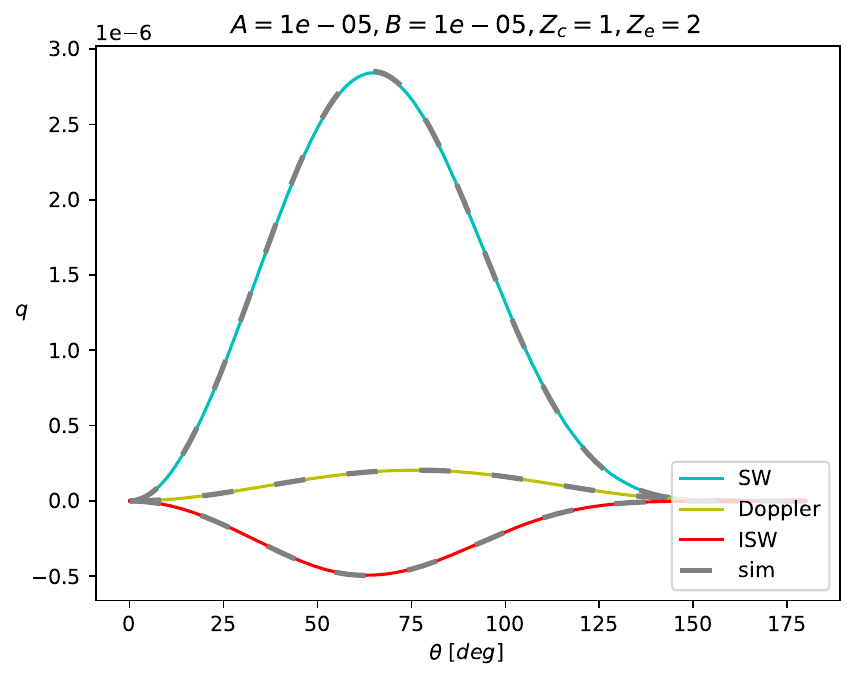}
  \caption{The RQF field distribution induced by the bubble collision contributed by SW, Doppler and ISW effects from the analytic calculation (solid lines) and from numerical calculation generated by \texttt{RemoteField} (dashed lines) with $Z_c=1$, $Z_e=2$ and $A=10^{-5}, B=A=10^{-5}$. It shows an excellent agreement between the two sets of result. }
    \label{fig:compare}
  \end{figure*}

\begin{figure*}[t]
  \includegraphics[width=0.8\textwidth]{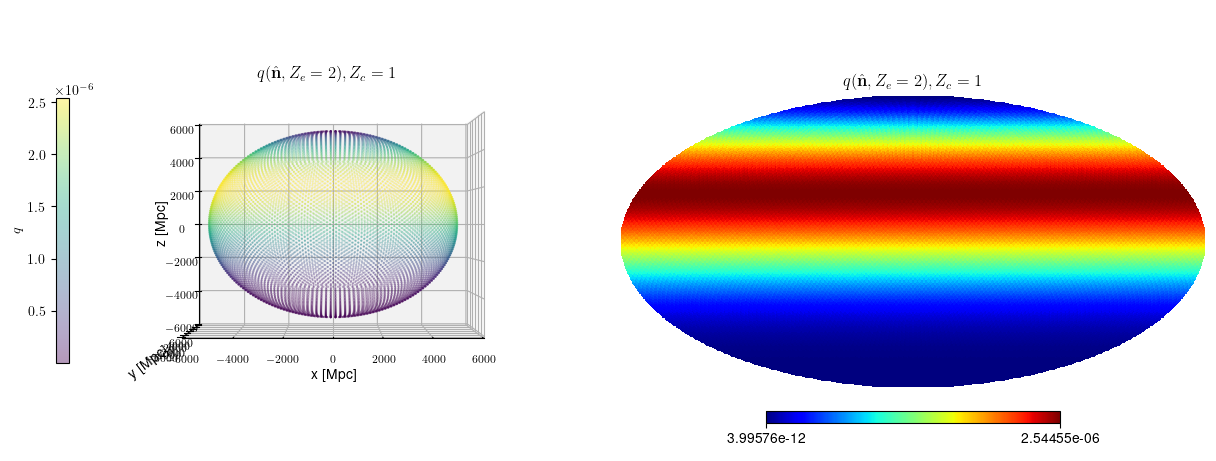}
  \caption{A numerically-calculated RQF induced by the bubble collision at $Z_e=2$ for $Z_c=1$ with $A=10^{-5}$ and $ B=10^{-5}$.
  The left panel shows the spherical distribution viewed from 3D perspective, and the right panel shows the corresponding Mollview map. They clearly show the azimuthal symmetric pattern.}
  \label{fig:RQF sphere}
\end{figure*}

In Fig.~\ref{fig:RQF zc=1} and Fig.~\ref{fig:RQF zc=2}, we show the 1D angular distribution of the RQF induced by the bubble collision, plotted as a function of the polar angle $\theta_e$. The panels, for $Z_e = 0, 1, 2, 4$, illustrate how the RQF signal evolves with the observer's position. The overall amplitude decays with increasing $Z_e$, as it scales with the line-of-sight distance to decoupling, $\Delta \chi_{\mathrm{dec}}$. As the electron's LSS shrinks at higher redshifts, it provides greater angular resolution on the collision boundary, causing the peak of the distribution to shift towards the northern hemisphere ($\theta_e < 90^\circ$), a feature that enhances the constraining power on its location. Furthermore, the quadratic potential ($B=1, A=0$) produces a signal more concentrated towards the pole compared to the linear potential ($A=1, B=0$), demonstrating the RQF's ability to differentiate between the two fundamental parameters. A comparison of the two figures also reveals that increasing the boundary distance $Z_c$ from 1 to 2 suppresses the amplitude from the quadratic term, while leaving the linear signal largely unaffected \footnote{We also note that there are no non-differentiable points in the RQF as those in the RDF for certain redshifts \cite{Zhang:2015uta}. This is because, unlike the RDF, the local Doppler effect does not contribute to the RQF, and hence the RQF distribution is smooth as a pure convolution over the space}.

To validate the analytic expressions for $q_{\alpha}$ (where $\alpha \in \{ \mathrm{SW, ISW, Doppler}\})$ derived above, we compare them with the numerical RQF results. These numerical results are generated by our \texttt{RemoteField} code using Eq.~\eqref{eq:RDF/RQF Fourier}.  As shown in Fig.~\ref{fig:compare}, the analytic and numerical calculations are in excellent agreement. To further illustrate the expected signal, Fig.~\ref{fig:RQF sphere} shows the full-sky numerical distribution of the RQF for $Z_c=1$ and $Z_e=2$, which clearly shows the azimuthally symmetric pattern. We have also numerically calculated the RDF and confirmed its agreement with the analytic result from \cite{Zhang:2015uta}. Our \texttt{RemoteField} code is not limited to bubble collision. It can be broadly applied to compute the RDF/RQF signals induced by other superhorizon-scale phenomena, such as cosmic topology or phase transition domain walls. This is particularly useful for scenarios lacking spatial symmetry where analytic solutions may be intractable,  but numerical approach remains viable. We leave the exploration of these other applications for future work.

\begin{figure}
     \centering
     \begin{subfigure}[b]{0.45\textwidth}
         \centering
         \includegraphics[width=\textwidth]{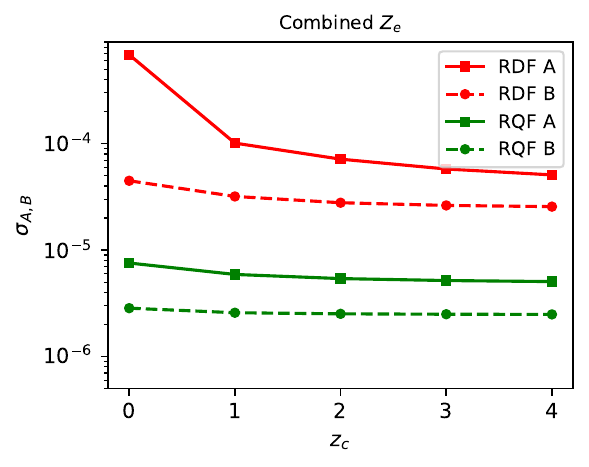}
         \caption{}
         \label{fig:single FullCov combined}
     \end{subfigure}
     \hfill
     \begin{subfigure}[b]{0.5\textwidth}
       \hspace*{-1.2cm}
         \centering
         \includegraphics[width=\textwidth]{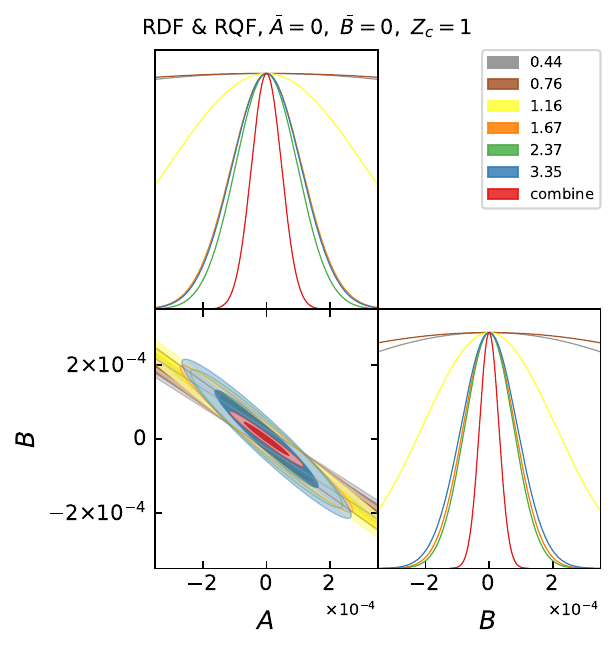}
         \caption{}
         \label{fig:countour FullCov}
       \end{subfigure}
       \caption{The forecast constraint on the bubble collision parameters using the $\ell=1$ mode of the RDF and the $\ell=2$ mode of the RQF E-mode with the full covariance including contributions from both reconstruction noise and $\Lambda \text{CDM}$ variance. (a) shows the 1$\sigma$ errors using each of the two modes with $Z_c= \{0, 1, 2, 3, 4\}$ for the case of $\bar{A}=0, B\neq 0$ and $A\neq 0, \bar{B}=0$ . All $Z_e$ bins are included here.
         (b) displays the joint 1$\sigma$ contours of the two parameters using each $Z_e$ bins (labeled by the central redshifts) and all the bins combined with $\bar{A}=0, \bar{B}=0$ and $Z_c=1$.}
       \label{fig:FullCov}
      \end{figure}

\begin{figure}
     \centering
     \begin{subfigure}[b]{0.45\textwidth}
         \centering
         \includegraphics[width=\textwidth]{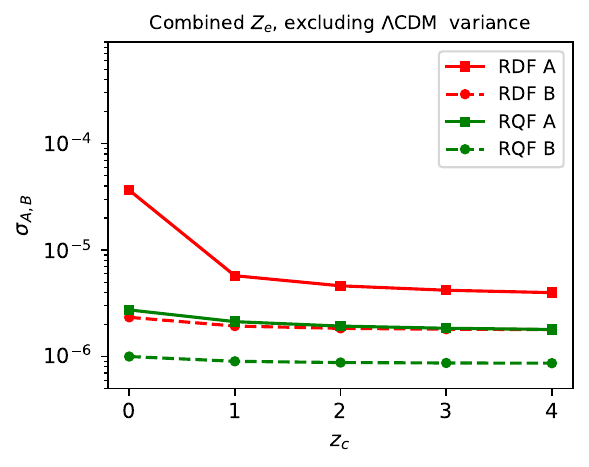}
         \caption{}
         \label{fig:single RecCov combined}
     \end{subfigure}
     \hfill
     \begin{subfigure}[b]{0.5\textwidth}
       \hspace*{-1.2cm}
         \centering
         \includegraphics[width=\textwidth]{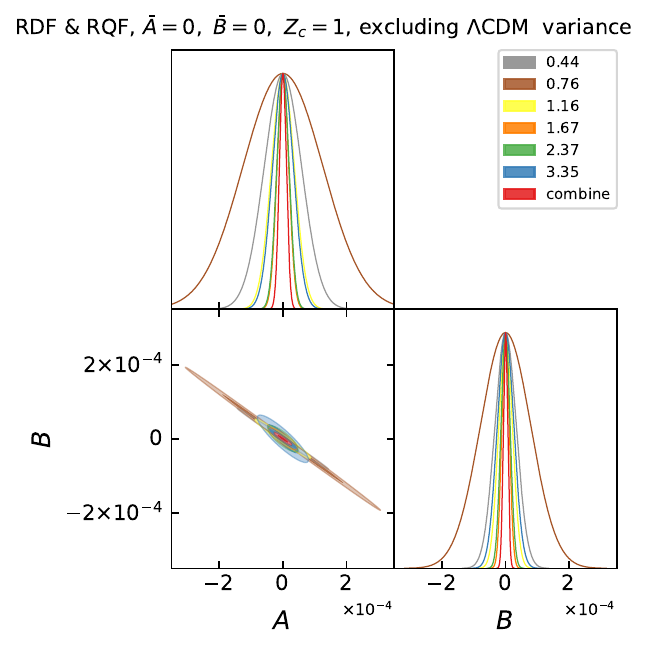}
         \caption{}
         \label{fig:countour RecCov}
     \end{subfigure}
     \caption{Same as Fig.~\ref{fig:FullCov} but with the variance from $\Lambda$CDM mitigated, i.e., only including the RDF/RQF reconstruction noise. This is a highly ideal estimation considering a complete $\Lambda \text{CDM}$ signal cleaning by RDF/RQF tomography.}
\end{figure}

\section{Forecast with RDF/RQF reconstruction}
\label{sec:forecast}
In this section, we forecast the expected constraints on the bubble collision parameters using the RDF/RQF reconstruction method within the Fisher matrix formalism. We first review the RDF/RQF reconstruction method in
\ref{subsec:RDF/RQF reconstruction} and then present the forecast results in \ref{subsec:Forecast}.

\subsection{Remote dipole and quadrupole fields reconstruction}
\label{subsec:RDF/RQF reconstruction}
The RDF/RQF can be reconstructed by combining the kSZ/pSZ signals in the CMB with galaxy surveys \cite{Deutsch:2018}. This method, analogous to CMB lensing reconstruction, leverages the fact that kSZ/pSZ effects introduce extra statistical anisotropies in CMB, modulated by the RDF/RQF and electron density fluctuations. For a given redshift bin labeled by $i$, the minimum variance estimators for the average RDF and RQF E-mode are constructed as \cite{Deutsch:2018}
\begin{equation}
  \label{eq:v and qe estimator}
\begin{aligned}
\hat{v}_{\ell m}^{i}= & N_{\ell}^{v, i} \sum_{\ell_1 m_1 \ell_2 m_2} (-1)^m\begin{pmatrix}
\ell & \ell_1 & \ell_2 \\
m & m_1 & m_2
\end{pmatrix} \\
                      & \times f^{v,i}_{\ell \ell_1 \ell_2} \frac{\Theta_{\ell_1 m_1}}{\h{C}_{\ell_1}^{TT}}\frac{\delta^{g^i}_{\ell_2 m_2}}{\h{C}_{\ell_2}^{g^ig^i}},\\
  \hat{q}_{\ell m}^{E, i}= & N_{\ell}^{q^E, i} \sum_{X=E,B} \sum_{\ell_1 m_1 \ell_2 m_2} \begin{pmatrix}
\ell & \ell_1 & \ell_2 \\
m & m_1 & m_2
\end{pmatrix} \\
                         & \times f^{q^{E,i}X}_{\ell \ell_1 \ell_2} \frac{X_{\ell_1, m_1}}{\h{C}^{XX}_{\ell1}}\frac{\delta^{g^i}_{\ell_2 m_2}}{\h{C}_{\ell_2}^{g^ig^i}},
\end{aligned}
\end{equation}
where $f^{v,i}_{\ell \ell_1 \ell_2}$ and $f^{q^{E,i}X}_{\ell \ell_1 \ell_2}$ (with $X \in \{E, B\}$ for CMB polarization) are the weight functions; $N_{\ell}^{v, i}$ and $N_{\ell}^{q^E, i}$ are the normalization factors of the estimators, which also represent the estimator variance or reconstruction noise; $\delta^{g^i}$ is the average galaxy density field in the redshift bin $i$; and the power spectra with hats in the denominators denote the observed power spectra, which include both signal and noise. The weight functions are given by
\begin{equation}
  \label{eq:weight functions}
  \begin{aligned}
    f^{v,i}_{\ell \ell_1 \ell_2} &=  \gamma_{\ell\ell_{1}\ell_{2}} \begin{pmatrix}
\ell & \ell_{1} & \ell_{2} \\
0 & 0 & 0
\end{pmatrix} C_{\ell_{2}}^{\tau^i g^i},\\
    f^{q^{E,i}X}_{\ell \ell_1 \ell_2} &=  -\frac{\sqrt{6}}{10}w_{\ell\ell_{1}\ell_{2}}^{q^{E} X} \gamma_{\ell\ell_{1}\ell_{2}}
\begin{pmatrix}
\ell & \ell_{1} & \ell_{2} \\
0 & 2 & -2
\end{pmatrix}
C_{\ell_{2}}^{\tau^i g^i},
  \end{aligned}
\end{equation}
with
\begin{equation}
  \label{eq:gamma}
\gamma_{\ell\ell_{1}\ell_{2}} = \sqrt{\frac{(2\ell+1)(2\ell_1+1)(2\ell_2+1)}{4\pi}},
\end{equation}
and $w_{\ell\ell_{1}\ell_{2}}^{q^{E} X}$ are the parity indicators for the RQF E-mode, defined as
\begin{equation}
  \label{eq:partity indices}
\begin{aligned}
  w_{\ell\ell_{1}\ell_{2}}^{q^{E} E} & = \frac{1}{2}[1 + (-1)^{\ell + \ell_1 + \ell_2}],\\
  w_{\ell\ell_{1}\ell_{2}}^{q^{E} B} & = \frac{i}{2}[1 - (-1)^{\ell + \ell_1 + \ell_2}].
\end{aligned}
\end{equation}
Here, $C_{\ell_{2}}^{\tau^i g^i}$ is the bin-averaged cross-correlation of the optical depth and galaxy field in redshift bin $i$ for a given model.

Several potential sources of systematic effects could bias the reconstruction. For the RDF reconstruction, these include optical depth bias due to mismatch between the true and assumed fiducial optical depth-galaxy cross-power, statistically isotropic CMB-galaxy cross-correlations, statistically anisotropic CMB-galaxy cross-correlations, and high-order noise bias (see \cite{Bloch:2024} for a review). Although a detailed investigation of systematics specifically for the RQF reconstruction is still lacking, they are expected to originate from similar sources.

\subsection{Forecast}
\label{subsec:Forecast}
We forecast the errors on the bubble collision parameters using the $\ell=1$ mode of the RDF and the $\ell=2$ mode of the RQF E-mode as observables in the Fisher matrix formalism. Following a similar approach as in \cite{Krywonos:2024mpb}, we model the likelihood function of the reconstructed multipoles, $\hat{v}^{i}_{1m}$ and $\hat{q}^i_{2m'}$, in a given redshift bin $i$ as
\begin{equation}
  \begin{aligned}
    &P(v^i_{1m}, q^i_{2m'} \mid \hat{v}^i_{1m}, \hat{q}^i_{2m'}) \propto\\
& \exp \Biggl[-f_{\mathrm{sky}}\biggl(\frac{3}{2(b^2_vC^{vv,i}_1 + N^{v,i}_1)} \sum_{m=-1}^1\left|\h{v}^i_{1 m} - v^i_{1 m}\right|^2 \\ &+ \frac{5}{2 (b^2_qC^{q^Eq^E, i}_2 + N^{q^E, i}_2)}\sum_{m'=-2}^2\left|\h{q}^i_{2 m'} - q^i_{2 m'} \right|^2\biggr)\Biggr] ,
    \end{aligned}
\end{equation}
where $f_{\mathrm{sky}}$ is the sky fraction. The total variance in the denominator of each term is the sum of the signal variance and the reconstruction noise. The signal variance is composed of the $\Lambda$CDM power spectra, $C^{vv,i}_1$ and $C^{q^Eq^E,i}_2$, and the optical depth bias factors, $b_v$ and $b_q$. These bias factors account for potential mismatches between the fiducial and true correlation between the optical depth and galaxy density fields \cite{Smith:2018, Cayuso:2021ljq, Giri:2020pkk, Bloch:2024}. While these prior studies have estimated $0.5 \lesssim b_v \lesssim 1.1$, the range of $b_q$ is less constrained. The $\Lambda$CDM power spectra are given by
  \begin{equation}
    \label{eq:RDF/RQF ps}
    \begin{aligned}
      C_L^{vv}(\chi_e) &=\int \frac{d k}{(2 \pi)^3}k^2 P_{\Psi}(k) \Delta_L^{v}(k, \chi_e)^2, \\
C_L^{q^Eq^E}(\chi_e) &=\int \frac{d k}{(2 \pi)^3}k^2 P_{\Psi}(k) \Delta_L^{q}(k, \chi_e)^2,
    \end{aligned}
  \end{equation}
which we compute for each redshift bin using \texttt{SZ\_cosmo}\footnote{\url{https://github.com/rcayuso/SZ_cosmo}}. The kernels $\Delta_L^{v}$ and $\Delta_L^{q}$ are detailed in Appendix~\ref{sec:appendix}. The reconstruction noise, $N^{v,i}_1$ and $N^{q^E,i}_2$, are calculated using the scale-independent approximation,
which is accurate in the limit of $\ell_1, \ell_2 \gg \ell$ \cite{Bloch:2024, Krywonos:2024mpb},
\begin{equation}
  \label{eq:norm}
  \begin{aligned}
    \frac{1}{N_{\ell}^{v, i}} &\approx  \frac{1}{N^{v, i}} \equiv \sum_{\ell'}\frac{2\ell' +1}{4\pi} \frac{(C^{\tau^i g^i}_{\ell'})^2}{\h{C}^{TT}_{\ell'}\h{C}^{g^ig^i}_{\ell'}}, \\
    \frac{1}{N_{\ell}^{q^E, i}} &\approx \frac{1}{N^{q^E, i}} \equiv \frac{6}{100}\sum_{X=E,B}\sum_{\ell'}\frac{2\ell' +1}{4\pi} \frac{(C^{\tau^i g^i}_{\ell'})^2}{\h{C}^{XX}_{\ell'}\h{C}^{g^ig^i}_{\ell'}},
  \end{aligned}
\end{equation}
where the hatted power spectra in the denominators, $\h{C}_{\ell}^{XX}$, represent the \emph{observed} power spectra, modeled as the sum of the theoretical $\Lambda$CDM signal, $C_{\ell}^{XX}$, and the instrumental noise, $N^{X}_{\ell}$ (i.e., $\h{C}_{\ell}^{XX} = C_{\ell}^{XX} + N^{X}_{\ell}$) with $X \in \{ {T, E, B, g^i}\}$. We use the best-fit cosmology from Planck 2018 \cite{Planck2018:VI:CP} for the $\Lambda$CDM power spectra and compute the bin-averaged cross-power spectrum of the optical depth and galaxy density, $C_{\ell}^{\tau^i g^i}$, using \texttt{ReCCO}\footnote{\url{https://github.com/jcayuso/ReCCO}}.

For our forecast we consider a CMB-S4-like experiment and an LSST-like galaxy survey. We model the CMB instrumental noise as
\begin{equation}
  \label{eq:CMB noise}
  \begin{aligned}
    N^T_{\ell}&= (\Delta T)^2\text{exp}\Biggl[ \frac{\ell(\ell+1)\theta^2_{\text{FWHM}}}{8\text{ln}2} \Biggr], \\
    N^E_{\ell} &= N^B_{\ell} = 2N^T_{\ell},
  \end{aligned}
\end{equation}
with a noise level of $\Delta T = 1\mu K'$ and a beam full-width-half-maximum (FWHM) of $\theta_{\text{FWHM}} = 1.4'$.
We assume the galaxy survey noise is dominated by shot noise, with a noise power spectrum given by
\begin{equation}
  \label{eq:shot noise}
N^{g,i}_{\ell} = \frac{1}{N_{{\rm gal},i}} = \Biggl(  \int^{\chi^{\text{max}}_{i}}_{\chi^{\text{min}}_{i}} d\chi\, n(z[\chi])\Biggr)^{-1},
\end{equation}
where $N_{{\rm gal},i}$ is the number of galaxies per steradian in the redshift bin $i$. The galaxy number density, $n(z)$, is modeled as
\begin{equation}
  \label{eq:g number density}
n(z) = n_{\text{gal}} \frac{1}{2z_0} \left(\frac{z}{z_0}\right)^2 \text{exp}(-z/z_0),
\end{equation}
with $n_{\text{gal}} = 40\, \text{arcmin}^{-2}$ and $z_0=0.3$.

For our analysis, we make two simplifying assumptions. First, following \cite{Zhang:2015uta}, we assume the direction of the bubble collision is known \textit{a priori}.\footnote{For an unknown collision direction, the analysis would require marginalizing the likelihood over all possible directions on the sky (see, e.g., \cite{Krywonos:2024mpb}).} This allows us to align the z-axis of our coordinate system with the collision's axis of symmetry. Due to the azimuthal symmetry of the bubble collision signal in this frame, only the $m=0$ multipoles ($v_{10}$ and $q_{20}$) are non-zero. Second, we set the optical depth bias factors to unity ($b_v = b_q = 1$). Under these assumptions, the likelihood for a single redshift bin $i$ simplifies to
\begin{equation}
  \begin{aligned}
    &P(v^{i}_{10}, q^{i}_{20} \mid \hat{v}^{i}_{10}, \hat{q}^{i}_{20}) \propto\\
& \exp \Biggl[-f_{\mathrm{sky}}\biggl(\frac{3}{2(C^{vv,i}_1 + N^{v,i}_1)} \left|\h{v}^{i}_{1 0} - v^{i}_{1 0}\right|^2 \\ &+ \frac{5}{2 (C^{q^Eq^E,i}_2 + N^{q^E,i}_2)}\left|\h{q}^{i}_{2 0} - q^{i}_{2 0} \right|^2\biggr)\Biggr].
    \end{aligned}
  \end{equation}
The fisher matrix for the parameters $\{A, B\}$ in each redshift bin is then given by
\begin{equation}
  \label{eq:fisher}
  \begin{aligned}
    F^{i}_{\alpha \beta} &\equiv \left.\left\langle\frac{\partial \ln \mathcal{L}}{\partial \lambda_\alpha } \frac{\partial \ln \mathcal{L}}{\partial \lambda_\beta}\right\rangle\right|_{\left\{\lambda_\gamma\right\}=\left\{\bar{\lambda}_\gamma\right\}}\\
    & = f_{\text{sky}}\Biggl(\frac{3}{2(C^{vv, i}_1 + N^{v, i}_1)} \frac{\partial{v^i_{10}}}{\partial{\lambda_\alpha}} \frac{\partial{v^i_{10}}}{\partial{\lambda_\beta}} \\ & \, \, \, \, \, \, \, \, \, +  \frac{5}{2(C^{q^Eq^E,i}_2  + N^{q^E,i}_2)}\frac{\partial{q^i_{20}}}{\partial{\lambda_\alpha}} \frac{\partial{q^i_{20}}}{\partial{\lambda_\beta}} \Biggr)|_{\left\{\lambda_\gamma\right\}=\left\{\bar{\lambda}_\gamma\right\}},
  \end{aligned}
\end{equation}
where $\left\{\lambda_\gamma\right\}$ refer to the parameters to be constraint, i.e., A and B, and $\bar{\lambda}_\gamma$ represent their fiducial values, i.e., $\bar{A}$ and $\bar{B}$. For the forecast, we divide the redshift range between $0.3 \leq z \leq 4$ into six equally spaced, top-hat bins.

We test the constraining power of our method under two conditions. First, we present our baseline forecast, where the total variance includes contributions from both the standard $\Lambda$CDM signal and the reconstruction noise. In this context, we analyze the constraints on each bubble collision parameter individually and then their joint constraints. Second, we explore a more optimistic scenario where the $\Lambda$CDM signal contamination could be mitigated using tomographic techniques, leaving only the reconstruction noise in the variance.

\subsubsection{Baseline forecast}
We begin by assessing the individual constraining power of the RDF dipole and the RQF E-mode quadrupole on parameters $A$ and $B$, assuming the other parameter is zero. Figure~\ref{fig:single FullCov combined} shows the $1\sigma$ error on each parameter for various collision redshifts ($Z_c=\{0, 1,2, 3, 4\}$), combining information from all electron redshift ($Z_e$) bins. The results are highly promising when compared to existing limits from the primary CMB, which place the $1\sigma$ upper limit at the order of $10^{-4}$ \cite{Feeney:2012hj, Zhang:2015uta}. For a CMB-S4-like and LSST-like experiment, the RDF dipole reconstruction alone provides a comparable constraint. The RQF E-mode quadrupole is even more powerful, improving the constraint by approximately an order of magnitude, benefiting from the smaller intrinsic $\Lambda$CDM variance at $\ell=2$. This forecast is conservative, as it only uses the lowest multipole of the signal. A preliminary estimate indicates that the signal amplitudes at $\ell \approx 10$ are only an order of magnitude smaller for both RDF and RQF. Including these higher multipoles could tighten the constraints by another factor of a few. This result underscores the power of using small-scale CMB anisotropies to probe superhorizon physics, effectively bypassing the cosmic variance that limits large-scale primary CMB measurements \footnote{To constrain the bubble collision with a maximum significance, one should conduct a joint analysis with both the primary CMB and CMB remote fields in the future experiment as discussed in \cite{Cayuso:2019hen}}.

Next, we perform a joint constraint on both $A$ and $B$. Figure~\ref{fig:countour FullCov} shows the resulting $1\sigma$ confidence contours for a collision at $Z_c=1$, assuming fiducial values of $\bar{A}=0$ and $\bar{B}=0$. The contours from individual redshift bins exhibit distinct degeneracy directions. By combining information across all redshift bins, this degeneracy is effectively broken, leading to significantly tighter constraints. The final marginalized $1\sigma$ errors are approximately $7 \times 10^{-5}$ for $A$ and $5 \times 10^{-5}$ for $B$.

\subsubsection{Optimistic forecast with $\Lambda$CDM signal removed}
The ultimate sensitivity of the RDF/RQF reconstruction is limited by variance from both $\Lambda$CDM fluctuations and reconstruction noise. However, since the RDF and RQF can be reconstructed in multiple redshift bins, it may be possible to use this tomographic information to distinguish the primordial bubble collision signal from the standard $\Lambda$CDM background. This concept, inspired by similar ideas for constraining cosmic birefringence \cite{Namikawa:2023zux} using RQF tomography, could allow for the mitigation of the $\Lambda$CDM contribution to the variance.
To gauge the potential of such a technique, we perform a highly optimistic forecast assuming the $\Lambda$CDM variance can be perfectly removed, leaving only the reconstruction noise:
\begin{equation}
  \label{eq:noise_only_variance}
  \begin{aligned}
C^{vv, i}_1 + N^{v, i}_1 &\rightarrow N^{v, i}_1, \\
C^{q^Eq^E,i}_2  + N^{q^E,i}_2 &\rightarrow N^{q^E,i}_2.
  \end{aligned}
\end{equation}
As shown in Figures~\ref{fig:single RecCov combined} and \ref{fig:countour RecCov}, this idealized scenario leads to an improvement in the constraining power by a factor of a few to an order of magnitude. We emphasize that this is an optimistic best-case estimate. A realistic analysis would need to account for residual uncertainties from the cleaning process. Nevertheless, it highlights the significant potential for future improvements, and we leave a detailed investigation of such tomographic methods to future work.

\section{Conclusion}
\label{sec:conclusion}
The remote dipole and quadrupole fields (RDF/RQF) offer a powerful pathway to probe superhorizon physics, bypassing the cosmic variance limitations of the primary CMB. In this work, we explored the potential of RDF/RQF to constrain bubble collision predicted by the theory of eternal inflation. In this scenario, the eternal inflation forms distinct expanding bubbles (pocket universe) which might collide with each others. These collisions can leave distinct, azimuthally-symmetric inhomogeneities on superhorizon scales, providing a unique observational target.

To model this signature, we derived the first analytic expression of the RQF signal induced by a bubble collision, complementing prior work on the RDF.
Our calculation leveraged the inherent azimuthal symmetry of the signal and explicitly derived contributions from the Sachs-Wolfe, Doppler, and integrated Sachs-Wolfe effects. We validated this result against a new public software tool we developed, \texttt{RemoteField}, which numerically calculates the RDF/RQF for any given primordial potential and shows excellent agreement with our analytic derivation. This tool is designed for general use and can be applied to other superhorizon phenomena where analytic solutions may be intractable.
Using these signal models, we forecasted constraints on bubble collision parameters using the RDF/RQF reconstruction method, which combines data from next-generation CMB (e.g., CMB-S4) and galaxy surveys (e.g., LSST). This method's fidelity improves with experiment quality and has been successfully tested on real data \cite{Bloch:2024, McCarthy:2024nik}. Our analysis showed that the RDF dipole ($\ell=1$) reconstruction yields constraints comparable to the primary CMB alone, while the RQF E-mode quadrupole ($\ell=2$) improves upon them by an order of magnitude. By combining these two multipoles, we formed a joint constraint that is further narrowed by integrating information from different redshift bins, helping to break parameter degeneracies. We estimated that including higher multipoles (up to $\ell \approx 10$) could tighten constraints by another factor of a few.


Looking forward, we identified a promising path for enhancing sensitivity: using RDF/RQF tomography to mitigate the standard $\Lambda$CDM background contamination. An optimistic forecast suggests this could improve constraints by another factor of a few to an order of magnitude. While a detailed analysis including realistic foregrounds and systematics is a necessary next step, this work establishes a robust framework for searching for bubble collisions. We anticipate that this methodology can be extended to probe a broader range of superhorizon physics, including signatures from cosmic topology and domain walls.

\section*{Acknowledgments}
We thank Arthur Kosowsky, Toshiya Namikawa and Zhao Chen for inspiring discussion. HC acknowledges support from
the National Key R\&D Program of China (2023YFA1607800, 2023YFA1607801, 2020YFC2201602),
CMS-CSST-2021-A02, and the Fundamental Research Funds for
the Central Universities.
YG acknowledges supports from the University of Toronto's Eric and Wendy Schmidt AI in Science Postdoctoral Fellowship, a program of Schmidt Sciences.
The Dunlap Institute is funded through an endowment established by the David Dunlap family and the University of Toronto. This work uses open source software including \texttt{healpy} \cite{2019JOSS....4.1298Z}, \texttt{SZ\_cosmo, ReCCO} \cite{Cayuso:2021ljq} and \texttt{Class} \cite{software:class}.
\appendix
\section{RDF and RQF evolution}
\label{sec:appendix}
This appendix provides the detailed equations for the RDF and RQF evolution used in the main text, following the formalism of \cite{Terrana:2016xvc, Deutsch:2017cja, Deutsch:2018}. The potential growth function and the velocity growth function for long-wavelength modes are defined as
\begin{equation}
    \label{eq:growth functions}
    \Psi(\bm{r}, t)=D_{\Psi}(t)\Psi_i(\bm{r}),\, \,  \bm{v}(\bm{r}, t)=-D_{v}(t)\nabla\Psi_i(\bm{r}).
\end{equation}
The Fourier kernels for the RDF ($\mathcal{K}$) and RQF ($\mathcal{G}$), which appear in Eq.~\eqref{eq:RDF/RQF Fourier}, are given by:
\begin{equation}
  \label{eq:RDF kernels}
  \begin{aligned}
\mathcal{K}_{\mathrm{SW}}(k, \chi_e) & =3(2 D_{\Psi}(\chi_{\mathrm{dec}})-\frac{3}{2}) j_1(k \Delta \chi_{\mathrm{dec}}), \\
\mathcal{K}_{\mathrm{ISW}}(k, \chi_e) & = 6\int_{a_{\mathrm{dec}}}^{a_e} d a \frac{d D_{\Psi}}{d a} j_1(k \Delta \chi(a)), \\
    \mathcal{K}_{\text{Doppler}}(k, \chi_e) & = k D_v(\chi_{\mathrm{dec}})\left[j_0(k \Delta \chi_{\mathrm{dec}}) -2j_2(k \Delta \chi_{\mathrm{dec}}) \right]\\   &-kD_v(\chi),
\end{aligned}
\end{equation}
and
\begin{equation}
  \label{eq:RQF kernels}
  \begin{aligned}
\mathcal{G}_{\mathrm{SW}}(k, \chi_e) & =-4 \pi(2 D_{\Psi}(\chi_{\mathrm{dec}})-\frac{3}{2}) j_2(k \Delta \chi_{\mathrm{dec}}), \\
\mathcal{G}_{\mathrm{ISW}}(k, \chi_e) & =-8 \pi \int_{a_{\mathrm{dec}}}^{a_e} d a \frac{d D_{\Psi}}{d a} j_2(k \Delta \chi(a)), \\
\mathcal{G}_{\text {Doppler }}(k, \chi_e) & =\frac{4 \pi}{5} k D_v(\chi_{\mathrm{dec}})\left[3 j_3(k \Delta \chi_{\mathrm{dec}})-2 j_1(k \Delta \chi_{\mathrm{dec}})\right].
\end{aligned}
\end{equation}


Finally, the multipole coefficients for the remote velocity field ($\Delta_{\ell}^{v}$) and remote quadrupole field ($\Delta_{\ell}^{q}$), which are used to calculate the power spectra in Eq.~\eqref{eq:RDF/RQF ps}, are defined as
\begin{equation}
\begin{aligned}
\Delta_{\ell}^{v}(k, \chi) = & \frac{4 \pi i^{\ell}}{2 \ell+1} \mathcal{K}^{v}(k, \chi)\left[\ell j_{\ell-1}(k \chi)\right.\\
&\left.-(\ell+1) j_{\ell+1}(k \chi)\right] T(k),
\end{aligned}
\end{equation}
and
\begin{equation}
\begin{aligned}
\Delta_{\ell}^{q}(k, \chi) = & -5 i^{\ell} \sqrt{\frac{3}{8}} \sqrt{\frac{(\ell+2) !}{(\ell-2) !}} \frac{j_{\ell}(k \chi)}{(k \chi)^{2}} T(k) \\
& \times\left[\mathcal{G}_{\mathrm{SW}}(k, \chi)+\mathcal{G}_{\mathrm{ISW}}(k, \chi)+\mathcal{G}_{\text{Doppler}}(k, \chi)\right],
\end{aligned}
\end{equation}
where the transfer function, $T(k)$, accounts for the sub-horizon evolution of small-scale modes. While we treat it generically in this analysis, it is typically modeled using a fitting function (e.g., the BBKS formula \cite{Bardeen:1986}) that depends on the cosmological parameters.

\vspace{1em}

\section{Bubble collision parameters}
\label{sec:appendix b}
Following \cite{Wainwright:2014pta}, we consider a model of inflation defined by a scalar lagrangian with one metastable false vacuum and at least one true vacuum. The tunneling behavior between the false vacuum and the true vacuum is described by the Coleman de Luccia \cite{Coleman:1977py, Callan:1977pt, Coleman:1980aw} (CDL) instanton, which is a non-perturbative solution of the corresponding equation of motion. In this scenario, the phenomenological bubble collision parameters $A$ and $B$ are connected to the fundamental parameters of the theory in the Newtonian gauge (see, e.g., Fig. 1 of \cite{Zhang:2015uta}) as
\begin{equation}
  \label{eq:A and B}
\begin{aligned}
&A = \frac{2}{5} \sqrt{\frac{8\Omega_{k}^{\text{obs}}}{r_{\text{obs}}}}\frac{\delta \phi_{0}^{\text{coll}}}{M_{\text{pl}}} (1 - \cos \Delta x_{\text{sep}}), \\
&B = \frac{2}{15}\Omega_{k}^{\text{obs}} \sqrt{\frac{r_{\text{coll}}}{r_{\text{obs}}}}\frac{H_{I}^{\text{coll}}}{H_{I}^{\text{obs}}} (1 - \cos \Delta x_{\text{sep}})^{2},
\end{aligned}
\end{equation}
\\
where \text{obs} and \text{coll} label the quantities in the observation bubble and the collision bubble respectively; $\Omega_{k}^{\text{obs}}$ is the energy density in curvature in the observation universe, $\delta \phi_{0}^{\text{coll}}$ is the distance between the instanton endpoints connecting the false vacuum
to the collision bubble interior, $r_{\text{obs}}$ and $r_{\text{coll}}$ represent the corresponding tensor-to-scalar ratios, and $H_I^{\text{obs}}$ and $H_I^{\text{coll}}$ are the corresponding Hubble scales during inflation;  $\Delta x_{\text{sep}}$ is the initial proper distance between the bubbles in the
collision frame measured in terms of the false vacuum Hubble scale ($0<\Delta x_{\text{sep}}<\pi$).

\bibliography{remote_field, birefringence,lensing,cite}

@article{Coleman:1977py,
    author = "Coleman, Sidney R.",
    title = "{The Fate of the False Vacuum. 1. Semiclassical Theory}",
    reportNumber = "HUTP-77-A004",
    doi = "10.1103/PhysRevD.16.1248",
    journal = "Phys. Rev. D",
    volume = "15",
    pages = "2929--2936",
    year = "1977",
    note = "[Erratum: Phys.Rev.D 16, 1248 (1977)]"
}

@article{Callan:1977pt,
    author = "Callan, Jr., Curtis G. and Coleman, Sidney R.",
    title = "{The Fate of the False Vacuum. 2. First Quantum Corrections}",
    reportNumber = "HUTP-77-A032",
    doi = "10.1103/PhysRevD.16.1762",
    journal = "Phys. Rev. D",
    volume = "16",
    pages = "1762--1768",
    year = "1977"
}

@article{Coleman:1980aw,
    author = "Coleman, Sidney R. and De Luccia, Frank",
    title = "{Gravitational Effects on and of Vacuum Decay}",
    reportNumber = "SLAC-PUB-2463",
    doi = "10.1103/PhysRevD.21.3305",
    journal = "Phys. Rev. D",
    volume = "21",
    pages = "3305",
    year = "1980"
}

@ARTICLE{2019JOSS....4.1298Z,
	author = {{Zonca}, Andrea and {Singer}, Leo and {Lenz}, Daniel and {Reinecke}, Martin and {Rosset}, Cyrille and {Hivon}, Eric and {Gorski}, Krzysztof},
	title = "{healpy: equal area pixelization and spherical harmonics transforms for data on the sphere in Python}",
	journal = {The Journal of Open Source Software},
	year = 2019,
	month = mar,
	volume = {4},
	number = {35},
	eid = {1298},
	pages = {1298},
	doi = {10.21105/joss.01298},
	adsurl = {https://ui.adsabs.harvard.edu/abs/2019JOSS....4.1298Z}
}

@article{Planck2018:VI:CP,
	archivePrefix = {arXiv},
	arxivId = {1807.06209},
	author = {{Planck Collaboration} and Aghanim, N. and Akrami, Y. and Ashdown, M. and Aumont, J. and others},
	eprint = {1807.06209},
	title = {{Planck 2018 results. VI. Cosmological parameters}},
	url = {http://arxiv.org/abs/1807.06209},
	year = {2018}
}

@article{software:class,
	archivePrefix = {arXiv},
	arxivId = {1104.2932},
	author = {Lesgourgues, Julien},
	eprint = {1104.2932},
	title = {{The Cosmic Linear Anisotropy Solving System (CLASS) I: Overview}},
	url = {http://arxiv.org/abs/1104.2932},
	year = {2011}
}

@article{ACT:2025fju,
    author = "Louis, Thibaut and others",
    collaboration = "ACT",
    title = "{The Atacama Cosmology Telescope: DR6 Power Spectra, Likelihoods and $Λ$CDM Parameters}",
    eprint = "2503.14452",
    archivePrefix = "arXiv",
    primaryClass = "astro-ph.CO",
    reportNumber = "FERMILAB-PUB-25-0071-PPD",
    month = "3",
    year = "2025"
}

@article{SPT-3G:2025bzu,
    author = "Camphuis, E. and others",
    collaboration = "SPT-3G",
    title = "{SPT-3G D1: CMB temperature and polarization power spectra and cosmology from 2019 and 2020 observations of the SPT-3G Main field}",
    eprint = "2506.20707",
    archivePrefix = "arXiv",
    primaryClass = "astro-ph.CO",
    reportNumber = "FERMILAB-PUB-25-0144-PPD",
    month = "6",
    year = "2025"
}

@article{Chiang:2009xsa,
    author = "Chiang, H. C. and others",
    title = "{Measurement of CMB Polarization Power Spectra from Two Years of BICEP Data}",
    eprint = "0906.1181",
    archivePrefix = "arXiv",
    primaryClass = "astro-ph.CO",
    doi = "10.1088/0004-637X/711/2/1123",
    journal = "Astrophys. J.",
    volume = "711",
    pages = "1123--1140",
    year = "2010"
}

@ARTICLE{Bardeen:1986,
       author = {{Bardeen}, J.~M. and {Bond}, J.~R. and {Kaiser}, N. and {Szalay}, A.~S.},
        title = "{The Statistics of Peaks of Gaussian Random Fields}",
      journal = {\apj},
         year = 1986,
        month = may,
       volume = {304},
        pages = {15},
          doi = {10.1086/164143},
       adsurl = {https://ui.adsabs.harvard.edu/abs/1986ApJ...304...15B}
}

@article{PhysRevLett.107.041301,
  title = {Confirmation of the Copernican Principle at Gpc Radial Scale and above from the Kinetic Sunyaev-Zel'dovich Effect Power Spectrum},
  author = {Zhang, Pengjie and Stebbins, Albert},
  journal = {Phys. Rev. Lett.},
  volume = {107},
  issue = {4},
  pages = {041301},
  numpages = {4},
  year = {2011},
  month = {Jul},
  publisher = {American Physical Society},
  doi = {10.1103/PhysRevLett.107.041301},
  url = {https://link.aps.org/doi/10.1103/PhysRevLett.107.041301}
}

@article{Feeney:2012hj,
    author = "Feeney, Stephen M. and Johnson, Matthew C. and McEwen, Jason D. and Mortlock, Daniel J. and Peiris, Hiranya V.",
    title = "{Hierarchical Bayesian Detection Algorithm for Early-Universe Relics in the Cosmic Microwave Background}",
    eprint = "1210.2725",
    archivePrefix = "arXiv",
    primaryClass = "astro-ph.CO",
    doi = "10.1103/PhysRevD.88.043012",
    journal = "Phys. Rev. D",
    volume = "88",
    pages = "043012",
    year = "2013"
}

@article{Meerburg:2017xga,
    author = "Meerburg, P. Daniel and Meyers, Joel and van Engelen, Alexander",
    title = "{Reconstructing the Primary CMB Dipole}",
    eprint = "1704.00718",
    archivePrefix = "arXiv",
    primaryClass = "astro-ph.CO",
    doi = "10.1103/PhysRevD.96.083519",
    journal = "Phys. Rev. D",
    volume = "96",
    number = "8",
    pages = "083519",
    year = "2017"
}

@article{Sunyaev:1980nv,
    author = "Sunyaev, R. A. and Zeldovich, Ya. B.",
    title = "{The Velocity of clusters of galaxies relative to the microwave background. The Possibility of its measurement}",
    journal = "Mon. Not. Roy. Astron. Soc.",
    volume = "190",
    pages = "413--420",
    year = "1980"
}

@article{Wainwright:2014pta,
    author = "Wainwright, Carroll L. and Johnson, Matthew C. and Aguirre, Anthony and Peiris, Hiranya V.",
    title = "{Simulating the universe(s) II: phenomenology of cosmic bubble collisions in full General Relativity}",
    eprint = "1407.2950",
    archivePrefix = "arXiv",
    primaryClass = "hep-th",
    doi = "10.1088/1475-7516/2014/10/024",
    journal = "JCAP",
    volume = "10",
    pages = "024",
    year = "2014"
}

@article{Krywonos:2024mpb,
    author = "Krywonos, Jordan and Hotinli, Selim C. and Johnson, Matthew C.",
    title = "{Constraints on cosmology beyond $\Lambda$CDM with kinetic Sunyaev Zel'dovich velocity reconstruction}",
    eprint = "2408.05264",
    archivePrefix = "arXiv",
    primaryClass = "astro-ph.CO",
    month = "8",
    year = "2024"
}

@article{Deutsch:2017cja,
    author = {Deutsch, Anne-Sylvie and Johnson, Matthew C. and M\"unchmeyer, Moritz and Terrana, Alexandra},
    title = "{Polarized Sunyaev Zel'dovich tomography}",
    eprint = "1705.08907",
    archivePrefix = "arXiv",
    primaryClass = "astro-ph.CO",
    reportNumber = "IGC-17-5-1",
    doi = "10.1088/1475-7516/2018/04/034",
    journal = "JCAP",
    volume = "04",
    pages = "034",
    year = "2018"
}

@article{Cayuso:2019hen,
    author = "Cayuso, Juan I. and Johnson, Matthew C.",
    title = "{Towards testing CMB anomalies using the kinetic and polarized Sunyaev-Zel\textquoteright{}dovich effects}",
    eprint = "1904.10981",
    archivePrefix = "arXiv",
    primaryClass = "astro-ph.CO",
    doi = "10.1103/PhysRevD.101.123508",
    journal = "Phys. Rev. D",
    volume = "101",
    number = "12",
    pages = "123508",
    year = "2020"
}

@article{Terrana:2016xvc,
    author = "Terrana, Alexandra and Harris, Mary-Jean and Johnson, Matthew C.",
    title = "{Analyzing the cosmic variance limit of remote dipole measurements of the cosmic microwave background using the large-scale kinetic Sunyaev Zel'dovich effect}",
    eprint = "1610.06919",
    archivePrefix = "arXiv",
    primaryClass = "astro-ph.CO",
    doi = "10.1088/1475-7516/2017/02/040",
    journal = "JCAP",
    volume = "02",
    pages = "040",
    year = "2017"
}

@ARTICLE{Smith:2018,
       author = {{Smith}, Kendrick M. and {Madhavacheril}, Mathew S. and {M{\"u}nchmeyer}, Moritz and {Ferraro}, Simone and {Giri}, Utkarsh and {Johnson}, Matthew C.},
        title = "{KSZ tomography and the bispectrum}",
      journal = {arXiv e-prints},
         year = 2018,
        month = oct,
          eid = {arXiv:1810.13423},
        pages = {arXiv:1810.13423},
          doi = {10.48550/arXiv.1810.13423},
archivePrefix = {arXiv},
       eprint = {1810.13423},
 primaryClass = {astro-ph.CO},
       adsurl = {https://ui.adsabs.harvard.edu/abs/2018arXiv181013423S}
}

@ARTICLE{Deutsch:2018,
       author = {{Deutsch}, Anne-Sylvie and {Dimastrogiovanni}, Emanuela and {Johnson}, Matthew C. and {M{\"u}nchmeyer}, Moritz and {Terrana}, Alexandra},
        title = "{Reconstruction of the remote dipole and quadrupole fields from the kinetic Sunyaev Zel'dovich and polarized Sunyaev Zel'dovich effects}",
      journal = {\prd},
         year = 2018,
        month = dec,
       volume = {98},
       number = {12},
          eid = {123501},
        pages = {123501},
          doi = {10.1103/PhysRevD.98.123501},
archivePrefix = {arXiv},
       eprint = {1707.08129},
 primaryClass = {astro-ph.CO},
       adsurl = {https://ui.adsabs.harvard.edu/abs/2018PhRvD..98l3501D}
}

@ARTICLE{Bloch:2024,
       author = {{Bloch}, Richard and {Johnson}, Matthew C.},
        title = "{Kinetic Sunyaev Zel'dovich velocity reconstruction from Planck and unWISE}",
      journal = {arXiv e-prints},
         year = 2024,
        month = may,
          eid = {arXiv:2405.00809},
        pages = {arXiv:2405.00809},
          doi = {10.48550/arXiv.2405.00809},
archivePrefix = {arXiv},
       eprint = {2405.00809},
 primaryClass = {astro-ph.CO},
       adsurl = {https://ui.adsabs.harvard.edu/abs/2024arXiv240500809B}
}

@article{Zhang:2015uta,
    author = "Zhang, Pengjie and Johnson, Matthew C.",
    title = "{Testing eternal inflation with the kinetic Sunyaev Zel'dovich effect}",
    eprint = "1501.00511",
    archivePrefix = "arXiv",
    primaryClass = "astro-ph.CO",
    doi = "10.1088/1475-7516/2015/06/046",
    journal = "JCAP",
    volume = "06",
    pages = "046",
    year = "2015"
}

@ARTICLE{1987ApJ...322..597V,
       author = {{Vishniac}, Ethan T.},
        title = "{Reionization and Small-Scale Fluctuations in the Microwave Background}",
      journal = {\apj},
         year = 1987,
        month = nov,
       volume = {322},
        pages = {597},
          doi = {10.1086/165755},
       adsurl = {https://ui.adsabs.harvard.edu/abs/1987ApJ...322..597V}
}

@article{Zhang:2003nr,
    author = "Zhang, Peng-Jie and Pen, Ue-Li and Trac, Hy",
    title = "{Precision era of the kinetic Sunyaev-Zeldovich effect: Simulations, analytical models and observations and the power to constrain reionization}",
    eprint = "astro-ph/0304534",
    archivePrefix = "arXiv",
    reportNumber = "FERMILAB-PUB-03-466-A",
    doi = "10.1111/j.1365-2966.2004.07298.x",
    journal = "Mon. Not. Roy. Astron. Soc.",
    volume = "347",
    pages = "1224",
    year = "2004"
}

@article{Sazonov:1999zp,
    author = "Sazonov, S. Y. and Sunyaev, R. A.",
    title = "{Microwave polarization in the direction of galaxy clusters induced by the CMB quadrupole anisotropy}",
    eprint = "astro-ph/9903287",
    archivePrefix = "arXiv",
    doi = "10.1046/j.1365-8711.1999.02981.x",
    journal = "Mon. Not. Roy. Astron. Soc.",
    volume = "310",
    pages = "765--772",
    year = "1999"
}

@article{Hand_2012,
   title={Evidence of Galaxy Cluster Motions with the Kinematic Sunyaev-Zel’dovich Effect},
   volume={109},
   ISSN={1079-7114},
   url={http://dx.doi.org/10.1103/PhysRevLett.109.041101},
   DOI={10.1103/physrevlett.109.041101},
   number={4},
   journal={Physical Review Letters},
   publisher={American Physical Society (APS)},
   author={Hand, Nick and Addison, Graeme E. and Aubourg, Eric and Battaglia, Nick and Battistelli, Elia S. and Bizyaev, Dmitry and Bond, J. Richard and Brewington, Howard and Brinkmann, Jon and Brown, Benjamin R. and Das, Sudeep and Dawson, Kyle S. and Devlin, Mark J. and Dunkley, Joanna and Dunner, Rolando and Eisenstein, Daniel J. and Fowler, Joseph W. and Gralla, Megan B. and Hajian, Amir and Halpern, Mark and Hilton, Matt and Hincks, Adam D. and Hlozek, Renée and Hughes, John P. and Infante, Leopoldo and Irwin, Kent D. and Kosowsky, Arthur and Lin, Yen-Ting and Malanushenko, Elena and Malanushenko, Viktor and Marriage, Tobias A. and Marsden, Danica and Menanteau, Felipe and Moodley, Kavilan and Niemack, Michael D. and Nolta, Michael R. and Oravetz, Daniel and Page, Lyman A. and Palanque-Delabrouille, Nathalie and Pan, Kaike and Reese, Erik D. and Schlegel, David J. and Schneider, Donald P. and Sehgal, Neelima and Shelden, Alaina and Sievers, Jon and Sifón, Cristóbal and Simmons, Audrey and Snedden, Stephanie and Spergel, David N. and Staggs, Suzanne T. and Swetz, Daniel S. and Switzer, Eric R. and Trac, Hy and Weaver, Benjamin A. and Wollack, Edward J. and Yeche, Christophe and Zunckel, Caroline},
   year={2012},
   month=jul }

@article{DeBernardis:2016pdv,
    author = "De Bernardis, F. and others",
    title = "{Detection of the pairwise kinematic Sunyaev-Zel'dovich effect with BOSS DR11 and the Atacama Cosmology Telescope}",
    eprint = "1607.02139",
    archivePrefix = "arXiv",
    primaryClass = "astro-ph.CO",
    doi = "10.1088/1475-7516/2017/03/008",
    journal = "JCAP",
    volume = "03",
    pages = "008",
    year = "2017"
}

@article{Chen:2021pwg,
    author = "Chen, Ziyang and Zhang, Pengjie and Yang, Xiaohu and Zheng, Yi",
    title = "{Detection of pairwise kSZ effect with DESI galaxy clusters and Planck}",
    eprint = "2109.04092",
    archivePrefix = "arXiv",
    primaryClass = "astro-ph.CO",
    doi = "10.1093/mnras/stab3604",
    journal = "Mon. Not. Roy. Astron. Soc.",
    volume = "510",
    number = "4",
    pages = "5916--5928",
    year = "2022"
}

@article{Kusiak:2021hai,
    author = "Kusiak, Aleksandra and Bolliet, Boris and Ferraro, Simone and Hill, J. Colin and Krolewski, Alex",
    title = "{Constraining the baryon abundance with the kinematic Sunyaev-Zel{\textquoteright}dovich effect: Projected-field detection using Planck, WMAP, and unWISE}",
    eprint = "2102.01068",
    archivePrefix = "arXiv",
    primaryClass = "astro-ph.CO",
    doi = "10.1103/PhysRevD.104.043518",
    journal = "Phys. Rev. D",
    volume = "104",
    number = "4",
    pages = "043518",
    year = "2021"
}

@article{Kamionkowski:1997na,
    author = "Kamionkowski, Marc and Loeb, Abraham",
    title = "{Getting around cosmic variance}",
    eprint = "astro-ph/9703118",
    archivePrefix = "arXiv",
    reportNumber = "CU-TP-822, CAL-632, CFA-4578",
    doi = "10.1103/PhysRevD.56.4511",
    journal = "Phys. Rev. D",
    volume = "56",
    pages = "4511--4513",
    year = "1997"
}

@article{McCarthy:2024nik,
    author = "McCarthy, Fiona and others",
    title = "{The Atacama Cosmology Telescope: Large-scale velocity reconstruction with the kinematic Sunyaev-Zel'dovich effect and DESI LRGs}",
    eprint = "2410.06229",
    archivePrefix = "arXiv",
    primaryClass = "astro-ph.CO",
    doi = "10.1088/1475-7516/2025/05/057",
    journal = "JCAP",
    volume = "05",
    pages = "057",
    year = "2025"
}

@article{Namikawa:2023zux,
    author = "Namikawa, Toshiya and Obata, Ippei",
    title = "{Cosmic birefringence tomography with polarized Sunyaev-Zel{\textquoteright}dovich effect}",
    eprint = "2306.08875",
    archivePrefix = "arXiv",
    primaryClass = "astro-ph.CO",
    doi = "10.1103/PhysRevD.108.083510",
    journal = "Phys. Rev. D",
    volume = "108",
    number = "8",
    pages = "083510",
    year = "2023"
}

@article{Schwarz:2015cma,
    author = "Schwarz, Dominik J. and Copi, Craig J. and Huterer, Dragan and Starkman, Glenn D.",
    title = "{CMB Anomalies after Planck}",
    eprint = "1510.07929",
    archivePrefix = "arXiv",
    primaryClass = "astro-ph.CO",
    doi = "10.1088/0264-9381/33/18/184001",
    journal = "Class. Quant. Grav.",
    volume = "33",
    number = "18",
    pages = "184001",
    year = "2016"
}

@article{Munchmeyer:2018eey,
    author = {M{\"u}nchmeyer, Moritz and Madhavacheril, Mathew S. and Ferraro, Simone and Johnson, Matthew C. and Smith, Kendrick M.},
    title = "{Constraining local non-Gaussianities with kinetic Sunyaev-Zel{\textquoteright}dovich tomography}",
    eprint = "1810.13424",
    archivePrefix = "arXiv",
    primaryClass = "astro-ph.CO",
    doi = "10.1103/PhysRevD.100.083508",
    journal = "Phys. Rev. D",
    volume = "100",
    number = "8",
    pages = "083508",
    year = "2019"
}

@article{Deutsch:2018umo,
    author = {Deutsch, Anne-Sylvie and Dimastrogiovanni, Emanuela and Fasiello, Matteo and Johnson, Matthew C. and M{\"u}nchmeyer, Moritz},
    title = "{Primordial gravitational wave phenomenology with polarized Sunyaev Zel{\textquoteright}dovich tomography}",
    eprint = "1810.09463",
    archivePrefix = "arXiv",
    primaryClass = "astro-ph.CO",
    doi = "10.1103/PhysRevD.100.083538",
    journal = "Phys. Rev. D",
    volume = "100",
    number = "8",
    pages = "083538",
    year = "2019"
}

@article{Garriga:2006hw,
    author = "Garriga, Jaume and Guth, Alan H. and Vilenkin, Alexander",
    title = "{Eternal inflation, bubble collisions, and the persistence of memory}",
    eprint = "hep-th/0612242",
    archivePrefix = "arXiv",
    reportNumber = "MIT-CTP-3800",
    doi = "10.1103/PhysRevD.76.123512",
    journal = "Phys. Rev. D",
    volume = "76",
    pages = "123512",
    year = "2007"
}

@article{Kosowsky:1991ua,
    author = "Kosowsky, Arthur and Turner, Michael S. and Watkins, Richard",
    title = "{Gravitational radiation from colliding vacuum bubbles}",
    reportNumber = "FERMILAB-PUB-91-323-A",
    doi = "10.1103/PhysRevD.45.4514",
    journal = "Phys. Rev. D",
    volume = "45",
    pages = "4514--4535",
    year = "1992"
}

@article{Erickcek:2008jp,
    author = "Erickcek, Adrienne L. and Carroll, Sean M. and Kamionkowski, Marc",
    title = "{Superhorizon Perturbations and the Cosmic Microwave Background}",
    eprint = "0808.1570",
    archivePrefix = "arXiv",
    primaryClass = "astro-ph",
    doi = "10.1103/PhysRevD.78.083012",
    journal = "Phys. Rev. D",
    volume = "78",
    pages = "083012",
    year = "2008"
}

@article{Cayuso:2021ljq,
    author = "Cayuso, Juan and Bloch, Richard and Hotinli, Selim C. and Johnson, Matthew C. and McCarthy, Fiona",
    title = "{Velocity reconstruction with the cosmic microwave background and galaxy surveys}",
    eprint = "2111.11526",
    archivePrefix = "arXiv",
    primaryClass = "astro-ph.CO",
    doi = "10.1088/1475-7516/2023/02/051",
    journal = "JCAP",
    volume = "02",
    pages = "051",
    year = "2023"
}

@article{Giri:2020pkk,
    author = "Giri, Utkarsh and Smith, Kendrick M.",
    title = "{Exploring KSZ velocity reconstruction with N-body simulations and the halo~model}",
    eprint = "2010.07193",
    archivePrefix = "arXiv",
    primaryClass = "astro-ph.CO",
    doi = "10.1088/1475-7516/2022/09/028",
    journal = "JCAP",
    volume = "09",
    pages = "028",
    year = "2022"
}
\end{document}